\documentclass[portuges,12pt,letter]{article}
\usepackage[centertags]{amsmath}
\usepackage{amsfonts}
\usepackage{newlfont}
\usepackage{amscd}
\usepackage{graphics}
\usepackage{epsfig}
\usepackage{indentfirst}
\usepackage{amsxtra}
\usepackage[latin1]{inputenc}
\usepackage{amssymb, amsmath}
\usepackage{amsthm}
\usepackage[mathscr]{eucal}

\newtheorem{thm}{Theorem}[section]

\newtheorem{obe}[thm]{Remark}
\newtheorem{remark}[thm]{Remark}
\author{Fabio Silva Botelho \\ Department of Mathematics \\  Federal University of Santa Catarina, UFSC \\
Florian\'{o}polis, SC - Brazil}

\title{\bf  A variational formulation for relativistic mechanics, a new interpretation for the Bohr atomic model and some concerning applications}
\date{}
\begin{document}
\maketitle

\abstract{This article develops a variational formulation for the relativistic Klein-Gordon equation.
The main results are obtained through an extension of the classical mechanics approach to a more general context, which  in some sense, includes the quantum mechanics one. For the second part of the text, the definition of  normal field and its relation with the wave function concept play a fundamental role in the main results establishment. Among the applications, we include  a model with the presence of electromagnetic fields and also the modeling of a chemical reaction. Finally, in the last section, we present some results about the Spin operator in a relativistic context.}

\section{Introduction}
In this work we propose a variational formulation  for the Klein-Gordon relativistic equation  obtained through an extension of the classical mechanics  approach to a more  general context.

 We introduce a energy part aiming to minimize and control, in a specific appropriate sense to be described in the next sections, the curvature field distribution along the concerned mechanical system.

 About the references, this work is based on the book \cite{500} and the articles \cite{46,48}. Indeed, in the next sections we present some  results similar to those  presented in \cite{500} and \cite{48}. In the third section we develop in details one of the  main results, namely, the establishment of the Klein-Gordon relativistic equation resulted from the  respective variational formulation.

At this point we remark that details on the Sobolev Spaces involved may be found in \cite{1,12}. For standard references in quantum mechanics, we refer to
\cite{55,1410,101} and the non-standard \cite{100}.

Finally, we emphasize this article is not about Bohmian mechanics, even though the David Bohm work has been always inspiring.

\section{The Newtonian approach}
In this section, specifically for a free particle context, we shall obtain a close relationship between classical and quantum mechanics.

Let $\Omega \subset \mathbb{R}^3$ be an open, bounded and connected set set with a regular (Lipschitzian) boundary denoted by $\partial \Omega$, on which we define a  position
field, in a free volume context, denoted by $\mathbf{r}: \Omega \times [0,T] \rightarrow \mathbb{R}^3$, where $[0,T]$ is a time interval.

Suppose also an associated density distribution scalar  field is given by $(\rho \circ \mathbf{r}): \Omega \times [0,T] \rightarrow [0,+\infty),$
so that the kinetics energy for such a system, denoted by $J:U \times V \rightarrow \mathbb{R}$, is defined as
$$J(\mathbf{r},\rho)=\frac{1}{2}\int_0^T\int_\Omega \rho(\mathbf{r}(\mathbf{x},t)) \frac{\partial \mathbf{r}(\mathbf{x},t)}{\partial t} \cdot \frac{\partial \mathbf{r}(\mathbf{x},t)}{\partial t} \sqrt{g}\;d\mathbf{x}dt,$$
subject to $$\int_\Omega \rho(\mathbf{r}(\mathbf{x},t)) \sqrt{g}\;d\mathbf{x}=m, \text{ on } [0,T],$$
where $m$ is the total system mass, $t$ denotes time and $d\mathbf{x}=dx_1\; dx_2\; dx_3.$

Here,
\begin{eqnarray}U&=&\{\mathbf{r} \in W^{1,2}(\Omega \times [0,T])\;:\; \mathbf{r}(\mathbf{x},0)=\mathbf{r}_0(\mathbf{x})
\nonumber \\ && \text{ and } \mathbf{r}(\mathbf{x},T)=\mathbf{r}_1(\mathbf{x}), \text{ in } \Omega\},
\end{eqnarray}
and
$$V=\{\rho(\mathbf{r}) \in L^2([0,T]; W^{1,2}(\Omega))\;:\; \mathbf{ r} \in U\}.$$
Also $$\mathbf{g}_k=\frac{\partial \mathbf{r}(\mathbf{x},t)}{\partial x_k},$$
where we assume $$\{\mathbf{g}_k,\; k \in \{1,2,3\}\}$$ to be a linearly independent set in $\Omega \times[0,T],$
$$g_{jk}=\mathbf{g}_j \cdot \mathbf{g}_k,$$  $$\{g^{ij}\}=\{g_{ij}\}^{-1},$$ and $$g=\det\{g_{jk}\}.$$

For such a standard Newtonian formulation, the kinetics energy takes into account just the tangential field given by the time derivative
$$\frac{\partial \mathbf{r}(\mathbf{x},t)}{\partial t}.$$

At this point, the  idea is to complement such an energy with a new term, denoted by $\hat{R}$, which would consider also the control of curvature distribution
along the mechanical system.

So, with such statements in mind, we redefine the concerning energy, denoting it again by $J:U \times V \times V_1 \rightarrow \mathbb{R}$, as
\begin{eqnarray}
J(\mathbf{r}, \rho)&=& -\frac{1}{2}\int_0^T\int_\Omega \rho(\mathbf{r}(\mathbf{x},t)) \frac{\partial \mathbf{r}(\mathbf{x},t)}{\partial t} \cdot \frac{\partial \mathbf{r}(\mathbf{x},t)}{\partial t} \sqrt{g}\;d\mathbf{x}dt \nonumber \\ &&+\frac{\gamma}{2}
\int_0^T \int_\Omega \hat{R} \sqrt{g}\;d\mathbf{x}dt,\end{eqnarray}
subject to
 $$\int_\Omega \rho(\mathbf{r}(\mathbf{x},t)) \sqrt{g}\;d\mathbf{x}=m, \text{ on } [0,T],$$
 where
$$\hat{R}=\sum_{i,j,k,l=1}^3\; g^{ij} g^{kl}\; \frac{\partial }{\partial x_i}\left( \sqrt{\frac{\rho(\mathbf{x},t)}{m}} \frac{\partial \mathbf{r}(\mathbf{x},t)}{\partial x_j}\right)
\cdot \frac{\partial }{\partial x_k}\left( \sqrt{\frac{\rho(\mathbf{x},t)}{m}} \frac{\partial \mathbf{r}(\mathbf{x},t)}{\partial x_l}\right),$$
and $\gamma>0$ is a constant to be specified.

Thus, defining a complex function $\phi$ such that
$$|\phi|= \sqrt{\frac{\rho}{m}}$$ and observing that the Christoffel symbols $\Gamma_{ij}^s$ are such that $$\frac{\partial^2 \mathbf{r}(\mathbf{x},t)}{\partial x_i \partial x_j}=
\sum_{s=1}^3\Gamma_{ij}^s \frac{\partial \mathbf{r}(\mathbf{x},t)}{\partial x_s},\; \forall i,j \in \{1,2,3\},$$
we have
\begin{eqnarray}
&&\frac{\partial}{\partial x_i}\left( \phi \frac{\partial \mathbf{r}(\mathbf{x},t)}{\partial x_j}\right)
\nonumber \\ &=& \frac{\partial \phi}{\partial x_i} \frac{\partial \mathbf{r}(\mathbf{x},t)}{\partial x_j}+\phi\; \frac{\partial^2 \mathbf{r}(\mathbf{x},t)}{\partial x_i \partial x_j}
\nonumber \\ &=& \frac{\partial \phi}{\partial x_i}\; \frac{\partial \mathbf{r}(\mathbf{x},t)}{\partial x_j}+\phi \;\sum_{s=1}^3\Gamma_{ij}^s \frac{\partial \mathbf{r}(\mathbf{x},t)}{\partial x_s}.
\end{eqnarray}
Therefore,
\begin{eqnarray}
&&\left(\frac{\partial}{\partial x_i}\left( \phi \frac{\partial \mathbf{r}(\mathbf{x},t)}{\partial x_j}\right)\right) \cdot
\left(\frac{\partial}{\partial x_k}\left( \phi^* \frac{\partial \mathbf{r}(\mathbf{x},t)}{\partial x_l}\right)\right)
\nonumber \\ &=& \left(\frac{\partial \phi}{\partial x_i} \frac{\partial \mathbf{r}(\mathbf{x},t)}{\partial x_j}+\phi \;\sum_{s=1}^3\Gamma_{ij}^s \;\frac{\partial \mathbf{r}(\mathbf{x},t)}{\partial x_s}\right)\cdot \left(\frac{\partial \phi^*}{\partial x_k} \frac{\partial \mathbf{r}(\mathbf{x},t)}{\partial x_l}+\phi^* \; \sum_{p=1}^3\Gamma_{kl}^p\; \frac{\partial \mathbf{r}(\mathbf{x},t)}{\partial x_p}\right)
\nonumber \\ &=& \sum_{s,p=1}^3\left(g_{jl} \;\frac{\partial \phi}{\partial x_i}\; \frac{\partial \phi^*}{\partial x_k} +\phi\; \frac{\partial \phi^*}{\partial x_k} \;\Gamma_{ij}^s \;g_{sl}\right.
\nonumber \\ &&\left.+ \phi^* \frac{\partial \phi}{\partial x_i}\; \Gamma_{kl}^p\;g_{pj}+|\phi|^2 \;\Gamma_{ij}^s \;\Gamma_{kl}^p \;g_{sp}\right).
\end{eqnarray}
From this, we may write,
\begin{eqnarray}
\hat{R}&=& \sum_{i,j,k,l,p,s=1}^3 \;g^{ij} g^{kl}\; \left(g_{jl}\; \frac{\partial \phi}{\partial x_i} \frac{\partial \phi^*}{\partial x_k} +\phi \;\frac{\partial \phi^*}{\partial x_k} \Gamma_{ij}^s \;g_{sl}\right.
\nonumber \\ &&+\left. \phi^*\; \frac{\partial \phi}{\partial x_i} \;\Gamma_{kl}^p\;g_{pj}+|\phi|^2\; \Gamma_{ij}^s\; \Gamma_{kl}^p \;g_{sp}\right).
\end{eqnarray}

Already including the Lagrange multipliers concerning the restrictions,  the final expression for the
energy, denoted by $J:U\times V  \rightarrow \mathbb{R}$, would be given by
\begin{eqnarray}
J(\mathbf{r}, \phi,E)&=& -\frac{1}{2}\int_0^T\int_\Omega m|\phi(\mathbf{r}(\mathbf{x},t))|^2 \frac{\partial \mathbf{r}(\mathbf{x},t)}{\partial t} \cdot \frac{\partial \mathbf{r}(\mathbf{x},t)}{\partial t} \sqrt{g}\;d\mathbf{x}dt \nonumber \\ &&+ \frac{\gamma}{2}
\int_0^T \int_\Omega \hat{R} \sqrt{g}\;d\mathbf{x}dt
\nonumber \\ &&- m\;\int_0^T E(t)\left( \int_\Omega |\phi(\mathbf{r})|^2\; \sqrt{g} \; d\mathbf{x}-1\right)\;dt,
\end{eqnarray}
where,
\begin{eqnarray}U&=&\{\mathbf{r} \in W^{1,2}(\Omega \times [0,T])\;:\; \mathbf{r}(\mathbf{x},0)=\mathbf{r}_0(\mathbf{x})
\nonumber \\ && \text{ and } \mathbf{r}(\mathbf{x},T)=\mathbf{r}_1(\mathbf{x}), \text{ in } \Omega\},
\end{eqnarray}

Finally, in particular for the special case in which
$$\mathbf{r}(\mathbf{x},t) \approx \mathbf{x},$$ we get $$\frac{\partial \mathbf{r}(\mathbf{x},t)}{\partial t}\approx 0,$$ $\mathbf{g}_k\approx \mathbf{e}_k$, where $$\{\mathbf{e}_1, \mathbf{e}_2 , \mathbf{e}_3\}$$ is the canonical basis of $\mathbb{R}^3.$

Therefore, in such a case,
$$ \frac{\gamma}{2}\int_0^T\int_\Omega \hat{R}\;\sqrt{g}\;d\mathbf{x} dt\approx \frac{\gamma T }{2} \sum_{k=1}^3\int_\Omega
\frac{\partial \phi}{\partial x_k} \frac{\partial \phi^*}{\partial x_k} \;d\mathbf{x}. $$

Hence, with such last results we may infer that
\begin{eqnarray}J(\mathbf{r}, \phi,E)/T&\approx& \tilde{J}(\phi, E)\nonumber \\ &=&\frac{\gamma  }{2} \sum_{k=1}^3\int_\Omega
\frac{\partial \phi}{\partial x_k} \frac{\partial \phi^*}{\partial x_k} \;d\mathbf{x} \nonumber \\ &&-E\left( \int_\Omega |\phi|^2d\mathbf{x}-1\right).
\end{eqnarray}

This last energy is just the standard Schr\"{o}dinger one in a free particle context.

\section{A brief note on the relativistic context, the Klein-Gordon equation}
 Of particular interest is the case in which  $\mathbf{x}=(x_1,x_2,x_3) \in \mathbb{R}^3$
and point-wise,
$$\mathbf{r}(\mathbf{x},t)=(ct,X_1(t,\mathbf{x}),X_2(t,\mathbf{x}),X_3(t,\mathbf{x})),$$
where
$$M=\{\mathbf{r}(\mathbf{x},t)\;:\; (\mathbf{x},t) \in \Omega \times [0,T]\},$$
for an appropriate $\Omega \subset \mathbb{R}^3.$

Also, denoting $d\mathbf{x}=dx_1dx_2dx_3,$ the mass differential would be given by $$dm=\frac{\rho(\mathbf{r})}{\sqrt{1-v^2/c^2}} \sqrt{-g}\;d\mathbf{x}= \frac{ |R(\mathbf{r})|^2}{\sqrt{1-v^2/c^2}}\sqrt{-g}\;d\mathbf{x},$$
and the semi-classical kinetics energy differential would be expressed by   \begin{eqnarray}dE_c&=&  \frac{\partial \mathbf{r}(t,\mathbf{x})}{\partial t} \cdot \frac{\partial \mathbf{r}(t,\mathbf{x})}{\partial t} \;dm\nonumber \\ &=& -\left(\frac{d\overline{t}}{dt}\right)^2\;dm \nonumber \\ &=&-(c^2-v^2)\;dm,\end{eqnarray} so that
$$dE_c= -c^2(\sqrt{1-v^2/c^2})|R(\mathbf{r})|^2 \sqrt{-g}\;d\mathbf{x},$$ where $$d\overline{t}^2=c^2dt^2-dX_1(t,\mathbf{x})^2-dX_2(t,\mathbf{x})^2-dX_3(t,\mathbf{x})^2.$$

Thus, the concerning energy is expressed by,
\begin{eqnarray}J_1(\mathbf{r},R)&=& -\int_0^T\int_\Omega dE_c \;dt+\frac{\gamma}{2}\int_0^T\int_\Omega \hat{R}\sqrt{-g}\;d\mathbf{x}\;dt \nonumber \\ &=&c^2\int_0^T\int_\Omega  |R(\mathbf{r})|^2 \sqrt{1-v^2/c^2}\;\sqrt{-g}\;d\mathbf{x}\;dt \nonumber \\ &&
+\frac{\gamma}{2}\int_0^T\int_\Omega \hat{R}\sqrt{-g}\;d\mathbf{x}\;dt,\end{eqnarray}
subject to
$$R(\mathbf{r}(\mathbf{x},0))=R_0(\mathbf{x})$$
$$R(\mathbf{r}(\mathbf{x},T))=R_1(\mathbf{x})$$
and $$R(\mathbf{r}(\mathbf{x},t))=0, \text{ on } \partial \Omega \times [0,T],$$

$$\int_\Omega  |R(\mathbf{r})|^2\;\sqrt{-g}\;d\mathbf{x} =m,
\text{ on } [0,T].$$

Here, we have denoted  $$x_0=ct,$$ $$(x_0,\mathbf{x})=(x_0,x_1,x_2,x_3),$$
$$\mathbf{g}_k=\frac{\partial\mathbf{r}(t,\mathbf{x})}{\partial x_k},$$
where we assume $$\{\mathbf{g}_k,\; k \in \{0,1,2,3\}\}$$ to be a linearly independent set in $\Omega \times[0,T],$
$$g=det\{g_{ij}\},$$ $$g_{ij}=\mathbf{g}_i \cdot \mathbf{g}_j,$$ $$\{g^{ij}\}=\{g_{ij}\}^{-1},$$
where such a  product is given by $$ \mathbf{y} \cdot \mathbf{z} =-y_0z_0+\sum_{i=1}^3y_i z_i, \; \forall \mathbf{y}=(y_0,y_1,y_2,y_3), \; \mathbf{z}=(z_0,z_1,z_2,z_3) \in \mathbb{R}^4.$$

Moreover,
$$\hat{R}=\sum_{i,j,k,l=0}^3\; g^{ij} g^{kl}\; \frac{\partial }{\partial x_i}\left( \frac{R(\mathbf{x},t)}{\sqrt{m}} \frac{\partial \mathbf{r}(\mathbf{x},t)}{\partial x_j}\right)
\cdot \frac{\partial }{\partial x_k}\left( \frac{R(\mathbf{x},t)}{\sqrt{m}} \frac{\partial \mathbf{r}(\mathbf{x},t)}{\partial x_l}\right).$$

Therefore, defining $\phi \in W^{1,2}(\Omega \times [0,T]; \mathbb{C})$ as  $$\phi(\mathbf{x},t)=\frac{R(\mathbf{r}(\mathbf{x},t))}{\sqrt{m}},$$
and recalling that the Christoffel symbols $\Gamma_{ij}^s$ are such that $$\frac{\partial^2 \mathbf{r}(\mathbf{x},t)}{\partial x_i \partial x_j}=
\sum_{s=0}^3\Gamma_{ij}^s \frac{\partial \mathbf{r}(\mathbf{x},t)}{\partial x_s},\; \forall i,j \in \{0,1,2,3\},$$
 similarly as in the last section, we may obtain
\begin{eqnarray}
\hat{R}&=& \sum_{i,j,k,l,p,s=0}^3\;g^{ij} g^{kl}\;\left(g_{jl}\; \frac{\partial \phi}{\partial x_i} \frac{\partial \phi^*}{\partial x_k} +\phi \;\frac{\partial \phi^*}{\partial x_k} \;\Gamma_{ij}^s \;g_{sl}\right.
\nonumber \\ &&+\left. \phi^*\; \frac{\partial \phi}{\partial x_i} \Gamma_{kl}^pg_{pj}+|\phi|^2\; \Gamma_{ij}^s\; \Gamma_{kl}^p\; g_{sp}\right).
\end{eqnarray}

Finally, we would also have $$v=\sqrt{\left(\frac{\partial X_1}{\partial t}\right)^2+\left(\frac{\partial X_2}{\partial t}\right)^2+\left(\frac{\partial X_3}{\partial t}\right)^2}.$$

In particular for the special case in which $$\mathbf{r}(\mathbf{x},t) \approx (ct,\mathbf{x}),$$ so that $$\frac{\partial \mathbf{r}(\mathbf{x},t)}{\partial t}\approx (c,0,0,0),$$  we would  obtain $$\mathbf{g}_0\approx (1,0,0,0),\; \mathbf{g}_1\approx (0,1,0,0),\;\mathbf{g}_2\approx (0,0,1,0) \text{ and } \mathbf{g}_3\approx (0,0,0,1) \in \mathbb{R}^4.$$
so that
\begin{eqnarray} \frac{\gamma}{2}\int_0^T\int_\Omega \hat{R}\;\sqrt{g}\;d\mathbf{x} dt &\approx& \frac{\gamma  }{2}\int_0^T\int_\Omega
\left(-\frac{1}{c^2} \frac{\partial \phi(\mathbf{x},t) }{\partial t} \frac{\partial \phi^*(\mathbf{x},t) }{\partial t}
 \right.\nonumber \\ &&\left.+ \sum_{k=1}^3\frac{\partial \phi(\mathbf{x},t)}{\partial x_k} \frac{\partial \phi^*(\mathbf{x},t)}{\partial x_k} \right)\;d\mathbf{x}dt, \end{eqnarray}
 and
 $$c^2\int_0^T\int_\Omega  |R(\mathbf{r})|^2 \sqrt{1-v^2/c^2}\;\sqrt{-g}\;d\mathbf{x}\;dt \approx m c^2\int_0^T\int_\Omega |\phi(\mathbf{x},t)|^2\;d\mathbf{x}dt.$$

Hence, with such last results we may infer that
\begin{eqnarray}J_1(\mathbf{r}, \phi,E)&\approx& \frac{\gamma}{2}\left(\int_0^T\int_\Omega
-\frac{1}{c^2} \frac{\partial \phi(\mathbf{x},t) }{\partial t} \frac{\partial \phi^*(\mathbf{x},t) }{\partial t}\;d\mathbf{x}dt
 \right.\nonumber \\ &&\left.+\sum_{k=1}^3\int_\Omega \int_0^T \frac{\partial \phi(\mathbf{x},t)}{\partial x_k} \frac{\partial \phi^*(\mathbf{x},t)}{\partial x_k}
 \;d\mathbf{x}dt\right) \nonumber \\ &&+ mc^2 \int_0^T\int_\Omega |\phi(\mathbf{x},t)|^2\;d\mathbf{x}dt  \nonumber \\ &&-m\int_0^TE(t)\left( \int_\Omega |\phi(\mathbf{x},t)|^2d\mathbf{x}-1\right)\;dt.
\end{eqnarray}

The Euler Lagrange equations for such an energy are given by

\begin{eqnarray}\label{US34}
&&\frac{\gamma}{2}\left(\frac{1}{c^2} \frac{\partial^2 \phi(\mathbf{x},t)}{\partial t^2}- \sum_{k=1}^3 \frac{\partial^2 \phi(\mathbf{x},t)}{\partial x_k^2} \right)
\nonumber \\ &&+mc^2\phi(\mathbf{x},t)- E_1(t) \phi(\mathbf{x},t)=0, \text{ in } \Omega,
\end{eqnarray}
where,
$$\phi(\mathbf{x},0)=\phi_0(\mathbf{x}), \text{ in } \Omega,$$
$$\phi(\mathbf{x},T)=\phi_1(\mathbf{x}), \text{ in } \Omega,$$
$$\phi(\mathbf{x},t)=0, \text{ on } \partial\Omega \times[0,T]$$
and $E_1(t)=mE(t).$

Equation (\ref{US34}) is the relativistic Klein-Gordon one.

For $E_1(t)=E_1 \in \mathbb{R}$ (not time dependent), at this point we suggest a solution (and implicitly related time boundary conditions) $\phi(\mathbf{x},t)=e^{-\frac{i E_1t}{\hbar}}\phi_2(\mathbf{x}),$ where $$\phi_2(\mathbf{x})=0, \text{ on } \partial \Omega.$$

Therefore, replacing this solution into equation (\ref{US34}), we would obtain
$$\left(\frac{\gamma}{2}\left(-\frac{E_1^2}{c^2\hbar^2}\phi_2(\mathbf{x}) -\sum_{k=1}^3 \frac{\partial^2 \phi_2(\mathbf{x})}{\partial x_k^2}\right)+mc^2\phi_2(\mathbf{x})- E_1 \phi_2(\mathbf{x})\right)e^{-\frac{i  E_1 t}{\hbar}}=0,$$
$\text{ in } \Omega.$

Denoting $$E_2=-\frac{\gamma  E_1^2}{2c^2 \hbar^2}+mc^2-E_1,$$ the final eigenvalue problem would stand for
$$-\frac{\gamma}{2} \sum_{k=1}^3 \frac{\partial^2 \phi_2(\mathbf{x})}{\partial x_k^2} +E_2\phi_2(\mathbf{x})=0, \text{ in } \Omega$$
where $E_1$ is such that
$$\int_\Omega |\phi_2(\mathbf{x})|^2\;d\mathbf{x}=1.$$

Moreover, from (\ref{US34}), such a solution $\phi(\mathbf{x},t)=e^{-\frac{i  E_1 t}{\hbar}}\phi_2(\mathbf{x})$ is also such that
\begin{eqnarray}\label{US35}
&&\frac{\gamma}{2}\left(\frac{1}{c^2} \frac{\partial^2 \phi(\mathbf{x},t)}{\partial t^2}- \sum_{k=1}^3 \frac{\partial^2 \phi(\mathbf{x},t)}{\partial x_k^2} \right)
\nonumber \\ &&+mc^2\phi(\mathbf{x},t)=i  \hbar\frac{\partial \phi(\mathbf{x},t)}{\partial t}, \text{ in } \Omega.
\end{eqnarray}
At this point, we recall that in quantum mechanics, $$\gamma=\hbar^2/m.$$

Finally, we remark this last equation (\ref{US35}) is a kind of relativistic Schr\"{o}dinger-Klein-Gordon equation.
\section{ A second model and the respective energy expression}

In a free volume context, denote again by $\mathbf{r}: \Omega \times [0,T] \rightarrow \mathbb{R}^3$  a position field, where $[0,T]$ is a time interval.

Suppose also  an associated density distribution scalar  field is given by $(\rho \circ \mathbf{r}): \Omega \times [0,T] \rightarrow [0,+\infty),$
so that the kinetics energy for such a system, denoted by $J:U \times V \rightarrow \mathbb{R}$, is defined as
$$J(\mathbf{r},\rho)=\frac{1}{2}\int_0^T\int_\Omega \rho(\mathbf{r}(\mathbf{x},t)) \frac{\partial \mathbf{r}(\mathbf{x},t)}{\partial t} \cdot \frac{\partial \mathbf{r}(\mathbf{x},t)}{\partial t} \sqrt{g}\;d\mathbf{x}dt,$$
subject to $$\int_\Omega \rho(\mathbf{r}(\mathbf{x},t)) \sqrt{g}\;d\mathbf{x}=m, \text{ on } [0,T].$$

We recall that for such a standard Newtonian formulation, the kinetics energy takes into account just the tangential field given by the time derivative
$$\frac{\partial \mathbf{r}(\mathbf{x},t)}{\partial t}.$$

Now the new idea is to complement such an energy with a new term which would consider also the variation of a normal field $\mathbf{n}$ and
concerning distribution of curvature, such that $$\mathbf{n} \cdot \frac{\partial \mathbf{r}(\mathbf{x},t)}{\partial t}=0, \text{ in }\Omega \times [0,T].$$

So, with such statements in mind, we redefine the concerning energy, denoting it again by $J:U \times V \times V_1 \rightarrow \mathbb{R}$, as
\begin{eqnarray}
J(\mathbf{r}, \mathbf{n}, \rho)&=& -\frac{1}{2}\int_0^T\int_\Omega \rho(\mathbf{r}(\mathbf{x},t)) \frac{\partial \mathbf{r}(\mathbf{x},t)}{\partial t} \cdot \frac{\partial \mathbf{r}(\mathbf{x},t)}{\partial t} \sqrt{g}\;d\mathbf{x}dt \nonumber \\ &&+
\frac{\gamma}{2}\int_0^T \int_\Omega \hat{R} \sqrt{g}\;d\mathbf{x}dt,\end{eqnarray}
where $\gamma>0$ is an appropriate constant,
%$$\mathbf{g}_k=\frac{\partial\mathbf{r}(\mathbf{x},t)}{\partial x_k},$$
%$$g=det\{g_{ij}\},$$ $$g_{ij}=\mathbf{g}_i \cdot \mathbf{g}_j,$$
$$\hat{R}=g^{ij}\hat{R}_{ij},$$
$$\hat{R}_{jk}=\hat{R}_{jik}^i,$$
$$\hat{R}_{jkl}^i=b^l_i\;b_{jk},$$
$$b_{ij}=-\frac{1}{\sqrt{m}}\frac{\partial \left(\sqrt{\rho(\mathbf{r})}\mathbf{n}(\mathbf{r})\right)}{\partial x_j}\cdot \mathbf{g}_i,$$
$$b_j^i=g^{il}b_{lj},$$
and,
$$\{g^{ij}\}=\{g_{ij}\}^{-1},$$
$\forall i,j,k,l \in \{1,2,3\}.$

subject to
$$ \mathbf{n}(\mathbf{r}) \cdot \mathbf{n}(\mathbf{r})  =1, \text{ in } \Omega \times [0,T],$$
 $$\mathbf{n}(\mathbf{r}) \cdot \frac{\partial \mathbf{r}}{\partial t}=0, \text{ in } \Omega \times [0,T],$$
and $$\int_\Omega \rho(\mathbf{r}(\mathbf{x},t)) \sqrt{g}\;d\mathbf{x}=m, \text{ on } [0,T].$$

Here $$V_1=\{\mathbf{n}(\mathbf{r}) \in L^2(\Omega \times [0,T])\;:\; \mathbf{r} \in U\}.$$

Thus, defining $\phi$ such that
$$|\phi|= \sqrt{\frac{\rho}{m}}$$ and already including the Lagrange multipliers concerning the restrictions,  the final expression for the
energy, denoted by $J:U\times V \times V_1 \times V_2 \times  [V_3]^2 \rightarrow \mathbb{R}$, would be given by
\begin{eqnarray}
J(\mathbf{r}, \mathbf{n}, \phi,E,\lambda_1,\lambda_2)&=& -\frac{1}{2}\int_0^T\int_\Omega m|\phi(\mathbf{r}(\mathbf{x},t))|^2 \frac{\partial \mathbf{r}(\mathbf{x},t)}{\partial t} \cdot \frac{\partial \mathbf{r}(\mathbf{x},t)}{\partial t} \sqrt{g}\;d\mathbf{x}dt \nonumber \\ &&+
\frac{\gamma}{2}\int_0^T \int_\Omega \hat{R} \sqrt{g}\;d\mathbf{x}dt
\nonumber \\ &&- m\;\int_0^T E(t)\left( \int_\Omega |\phi(\mathbf{r})|^2\; \sqrt{g} \; d\mathbf{x}-1\right)\;dt
\nonumber \\ &&+\langle \lambda_1, \mathbf{n} \cdot \mathbf{n}-1 \rangle_{L^2} \nonumber \\ &&+\left\langle \lambda_2, \mathbf{n} \cdot \frac{\partial \mathbf{r}}{\partial t} \right\rangle_{L^2},\end{eqnarray}
where,
\begin{eqnarray}U&=&\{\mathbf{r} \in W^{1,2}(\Omega \times [0,T])\;:\; \mathbf{r}(\mathbf{x},0)=\mathbf{r}_0(\mathbf{x})
\nonumber \\ && \text{ and } \mathbf{r}(\mathbf{x},T)=\mathbf{r}_1(\mathbf{x}), \text{ in } \Omega\},
\end{eqnarray}

$$V=\{\phi(\mathbf{r}) \in L^2([0,T]; W^{1,2}(\Omega;\mathbb{C}))\;:\; \mathbf{ r} \in U\},$$
$$V_1=\{\mathbf{n}(\mathbf{r}) \in L^2(\Omega \times [0,T])\;:\; \mathbf{r} \in U\},$$
$$V_2=L^2([0,T]),$$
 $$V_3=L^2(\Omega \times [0,T]).$$
 %and generically $$\langle f, h \rangle_{L^2}=\int_0^T\int_\Omega f h \sqrt{g} \;d\mathbf{x}\;dt, \forall f,h \in L^2(\Omega \times [0,T]).$$

Moreover,
%$$\mathbf{g}_k=\frac{\partial\mathbf{r}(\mathbf{x},t)}{\partial x_k},$$
%$$g=det\{g_{ij}\},$$ $$g_{ij}=\mathbf{g}_i \cdot \mathbf{g}_j,$$
$$\hat{R}=g^{ij}\hat{R}_{ij},$$
$$\hat{R}_{jk}=\hat{R}_{jik}^i,$$
$$\hat{R}_{jkl}^i=b^l_i\;b_{jk}^*,$$
$$b_{ij}=-\frac{\partial \left(\phi(\mathbf{r})\mathbf{n}(\mathbf{r})\right)}{\partial x_j}\cdot \mathbf{g}_i,$$
$$b_j^i=g^{il}b_{lj},$$
%and,
%$$\{g^{ij}\}=\{g_{ij}\}^{-1},$$
$\forall i,j,k,l \in \{1,2,3\}.$

Finally, in particular for the special case in which $$\mathbf{r}(\mathbf{x},t) \approx \mathbf{x},$$ so that $$\frac{\partial \mathbf{r}(\mathbf{x},t)}{\partial t}\approx 0,$$ and
$$\mathbf{n} \cdot \frac{\partial \mathbf{r}}{\partial t} \approx 0,$$ we may set $$\mathbf{n}=\mathbf{c},$$ where $\mathbf{c} \in \mathbb{R}^3$ is a constant
such that
$$\mathbf{c} \cdot \mathbf{c}=1,$$ and obtain $$\mathbf{g}_k \approx \mathbf{e}_k,$$
where $$\{\mathbf{e}_1, \mathbf{e}_2 , \mathbf{e}_3\}$$ is the canonical basis of $\mathbb{R}^3.$

Therefore, in such a case,
$$\frac{\gamma }{2} \int_0^T\int_\Omega \hat{R}\;\sqrt{g}\;d\mathbf{x} dt\approx \frac{\gamma T }{2} \sum_{k=1}^3\int_\Omega
\frac{\partial \phi}{\partial x_k} \frac{\partial \phi^*}{\partial x_k} \;d\mathbf{x}. $$

Hence, we would also obtain
\begin{eqnarray}J(\mathbf{r}, \mathbf{n}, \phi,E,\lambda_1,\lambda_2)/T&\approx& \tilde{J}(\phi, E)\nonumber \\ &=&\frac{\gamma  }{2} \sum_{k=1}^3\int_\Omega
\frac{\partial \phi}{\partial x_k} \frac{\partial \phi^*}{\partial x_k} \;d\mathbf{x} \nonumber \\ &&-E\left( \int_\Omega |\phi|^2d\mathbf{x}-1\right).
\end{eqnarray}

This last energy is just the standard Schr\"{o}dinger one in a free particle context.

\section{A brief note on the relativistic context for such a second model}
\noindent We recall to have denoted by $c$ the speed of light and $$d\overline{t}^2=c^2dt^2-dX_1^2-dX_2^2-dX_3^2.$$

In a relativistic free particle context,
 the Hilbert variational formulation could be extended, for a motion in a pseudo Riemannian relativistic $C^1$ class manifold $M$, where locally
 $$M=\{\mathbf{r}(\mathbf{u})\;:\; \mathbf{u} \in \Omega\},$$
  $$\mathbf{u}=(u_1,u_2,u_3,u_4) \in \mathbb{R}^4,$$ and $$\mathbf{r}:\Omega \subset \mathbb{R}^4 \rightarrow \mathbb{R}^4$$
  point-wise stands for,
  $$\mathbf{r}(\mathbf{u})=(ct(\mathbf{u}),X_1(\mathbf{u}),X_2(\mathbf{u}),X_3(\mathbf{u})),$$ to a functional $J_1$ where denoting
  $\rho(\mathbf{r})=|R(\mathbf{r})|^2$, the mass differential is given by $$dm=\frac{\rho(\mathbf{r})}{\sqrt{1-v^2/c^2}} \sqrt{|g|}\;d\mathbf{u}= \frac{ |R(\mathbf{r})|^2}{\sqrt{1-v^2/c^2}}\sqrt{|g|}\;d\mathbf{u},$$
 the semi-classical kinetics energy differential is given by   \begin{eqnarray}dE_c&=& \frac{\partial \mathbf{r}(\mathbf{u})}{\partial t}\cdot \frac{\partial \mathbf{r}(\mathbf{u})}{\partial t} \;dm\nonumber \\ &=& -\left(\frac{d\overline{t}}{dt}\right)^2\;dm \nonumber \\ &=&-(c^2-v^2)\;dm,\end{eqnarray} so that
$$dE_c= -c^2(\sqrt{1-v^2/c^2})|R(\mathbf{r})|^2 \sqrt{|g|}\;d\mathbf{u},$$ and

\begin{eqnarray}J_1(\mathbf{r},R,\mathbf{n})&=&- \int_\Omega dE_c+\frac{\gamma}{2}\int_\Omega \hat{R}\sqrt{|g|}\;d\mathbf{u} \nonumber \\ &=&c^2\int_\Omega  |R(\mathbf{r})|^2 \sqrt{1-v^2/c^2}\;\sqrt{|g|}\;d\mathbf{u} \nonumber \\ &&
+\frac{\gamma}{2}\int_\Omega \hat{R}\sqrt{|g|}\;d\mathbf{u},\end{eqnarray}
subject to
$$\int_\Omega |R(\mathbf{r})|^2\;\sqrt{|g|}\;d\mathbf{u} =m, $$
where $m$ is the particle mass at rest.

Moreover,

$$ \mathbf{n}(\mathbf{r})\cdot \frac{\partial \mathbf{r}}{\partial \overline{t}}  =0, \; \text{ in } \Omega,$$

where \begin{eqnarray}\frac{\partial \mathbf{r}}{\partial \overline{t}}&=& \frac{\partial \mathbf{r}}{\partial t}
\frac{\partial t}{\partial \overline{t}} \nonumber \\ &=& \frac{\frac{\partial \mathbf{r}}{\partial t}}{
\frac{\partial \overline{t}}{\partial t} } \nonumber \\ &=& \frac{\partial \mathbf{r}}{c\partial t}\frac{1}{\sqrt{1-v^2/c^2}},
\end{eqnarray}
and
$$ \mathbf{n}(\mathbf{r}) \cdot \mathbf{n}(\mathbf{r})  =1, \text{ in } \Omega.$$

Where  $\gamma$ is an appropriate positive constant to be specified.

Also,
$$\mathbf{g}_k=\frac{\partial\mathbf{r}(\mathbf{u})}{\partial u_k},$$
$$g=det\{g_{ij}\},$$ $$g_{ij}=\mathbf{g}_i \cdot \mathbf{g}_j,$$
where here,  in this subsection, such a  product is given by $$ \mathbf{y} \cdot \mathbf{z} =-y_0z_0+\sum_{i=1}^3y_i z_i, \; \forall \mathbf{y}=(y_0,y_1,y_2,y_3), \; \mathbf{z}=(z_0,z_1,z_2,z_3) \in \mathbb{R}^4,$$
$$\hat{R}=g^{ij}\hat{R}_{ij},$$
$$\hat{R}_{jk}=\hat{R}_{jik}^i,$$
$$\hat{R}_{jkl}^i=b^l_i\;b_{jk}^*,$$
$$b_{ij}=-\frac{1}{\sqrt{m}}\frac{\partial \left(R(\mathbf{r})\mathbf{n}(\mathbf{r})\right)}{\partial u_j}\cdot \mathbf{g}_i,$$
$$b_j^i=g^{il}b_{lj},$$
and,
$$\{g^{ij}\}=\{g_{ij}\}^{-1},$$
$\forall i,j,k,l \in \{1,2,3,4\}.$

Finally, $$v=\sqrt{\left(\frac{\partial X_1}{\partial t}\right)^2+\left(\frac{\partial X_2}{\partial t}\right)^2+\left(\frac{\partial X_3}{\partial t}\right)^2},$$

where, \begin{eqnarray} \frac{\partial X_k(\mathbf{u})}{\partial t}&=&\frac{\partial X_k(\mathbf{u})}{\partial u_j}\frac{\partial u_j}{\partial t}
\nonumber \\ &=&\sum_{j=1}^4\frac{\frac{\partial X_k(\mathbf{u})}{\partial u_j}}{\frac{\partial t(\mathbf{u})}{\partial u_j}},
\;\forall k \in \{1,2,3\}. \end{eqnarray}
Here the Einstein sum convention holds.
\begin{remark} The role of the variable $\mathbf{u}$ concerns the  idea of establishing a relation between $t,X_1,X_2$ and $X_3$. The dimension of $M$ may vary with the problem in question.
\end{remark}

\section{A brief note on the case including electro-magnetic effects}

In this section we address in a specific special relativistic context, the inclusion of electromagnetic effects.
\subsection{ About a specific Lorentz transformation}
In this section we assume the particle/volume motion is such that we are in a special relativity approximate context.

Consider the specific Lorentz transformation defined by a matrix $\{a_{jk}\}$, where the coordinates of the cartesian systems
$$(\mathbf{x},t)=(x_1,x_2,x_3,t)=(x_1,x_2,x_3,x_4/(i_mc))$$ and $$(\mathbf{x}',t')=(x_1',x_2',x_3',t')=(x_1',x_2',x_3',x_4'/(i_mc))$$ are related by the equations,
$$x_j'=\sum_{k=1}^4a_{jk}x_k,\; \forall j \in \{1,2,3,4\},$$
where $$x_4=i_mct\;\text{ and } x_4'=i_mct'.$$

More specifically, we consider the case in which $\{a_{jk}\}$ is generated by a motion of the origin $0'$ of the system $(\mathbf{x}',t')$  with velocity
$\mathbf{v}=(v_1,v_2,v_3)$ in relation to the origin $0$ of the system $(\mathbf{x},t).$ In such a  motion, we assume the axis $0'x_j'$ keeps parallel to
$0x_j$, $\forall j \in \{1,2,3\}.$

So, indeed, $\{a_{jk}\}$ is such that
$$x_j'=x_j+\left(\frac{1}{\sqrt{1-\frac{v^2}{c^2}}}-1\right)\frac{\mathbf{x}\cdot \mathbf{v}}{v^2}v_j-\frac{tv_j}{\sqrt{1-\frac{v^2}{c^2}}},$$
$\forall j \in \{1,2,3\},$
and
$$x_4'/(i_m c)=t'=\frac{1}{\sqrt{1-\frac{v^2}{c^2}}}\left(t-\frac{\mathbf{x}\cdot \mathbf{v}}{c^2}\right),$$
where, as above indicated
$$\mathbf{x}=(x_1,x_2,x_3),$$
$$\mathbf{x}'=(x_1',x_2',x_3'),$$
and,
$$\mathbf{v}=(v_1,v_2,v_3).$$

\subsection{Describing the self interaction energy and obtaining a final variational formulation}

Considering the model for an electronic field with position field given by $\mathbf{r}(\mathbf{x},t)$ over the set $\Omega \times [0,T],$ where $\Omega \subset \mathbb{R}^3$, and a mass/charge density given by $\rho(\mathbf{r})=|R(\mathbf{r})|^2$, we shall define the self interaction electric field differential, as indicated in the next lines.

First, denote in this section $dx=dx_1\;dx_2\;dx_3$, $$\mathbf{r}(x,t)=(ct,X_1(x,t),X_2(x,t),X_3(x,t)),$$ $$\mathbf{r}_1(x,t)=(X_1(x,t),X_2(x,t),X_3(x,t)),$$ $$\Delta \mathbf{r}_1(x,\tilde{x},t)=\mathbf{r}_1(x,t)-\mathbf{r}_1(\tilde{x},t),$$
and
$$dq(\tilde{x},t)=K_1 |R(\mathbf{r}(\tilde{x},t))|^2\sqrt{g(\mathbf{r}(\tilde{x},t))}d\tilde{x}.$$

Denote $$\mathbf{v}=\frac{\partial \mathbf{r}_1(\tilde{x},t)}{\partial t}-\frac{\partial \mathbf{r}_1(x,t)}{\partial t},$$

and define $$\Delta t(x,\tilde{x},t)=-\frac{1}{\sqrt{1-\frac{v^2}{c^2}}}\frac{\Delta \mathbf{r}_1(x,\tilde{x},t)) \cdot \mathbf{v}}{c^2}.$$

Define also $$\mathbf{v}'=\frac{\partial \mathbf{r}_1(\tilde{x},t-\Delta t(x,\tilde{x},t))}{\partial t}-\frac{\partial \mathbf{r}_1(x,t-\Delta t(x,\tilde{x},t))}{\partial t},$$
and,
\begin{eqnarray}\Delta \hat{\mathbf{r}}_1(x,\tilde{x},t)&=&\Delta \mathbf{r}_1(x,\tilde{x},t-\Delta t(x,\tilde{x},t))
\nonumber \\ &&+\left(\frac{1}{\sqrt{1-\frac{(v')^2}{c^2}}}-1\right)
\frac{(\Delta \mathbf{r}_1(x,\tilde{x},t-\Delta t(x,\tilde{x},t))  \cdot \mathbf{v}') \mathbf{v}'}{c^2},\end{eqnarray}
where $$\Delta \mathbf{r}_1(x,\tilde{x},t-\Delta t(x,\tilde{x},t))=\mathbf{r}_1(x,t-\Delta t(x,\tilde{x},t))-\mathbf{r}_1(\tilde{x},t-\Delta t(x,\tilde{x},t)).$$
Thus, the electric field generated by $dq(\tilde{x},t-\Delta t(x,\tilde{x},t))$ at $\mathbf{r}(x,t)$ is given by
\begin{eqnarray}d\mathbf{E}'(x,\tilde{x},t)&=&  K dq(\tilde{x},t-\Delta t(x,\tilde{x},t)) \frac{\Delta \hat{\mathbf{r}}_1(x,\tilde{x},t)}{|\Delta \hat{\mathbf{r}}_1(x,\tilde{x},t)|^3}
\nonumber \\ &=& K K_1 |R(\mathbf{r}(\tilde{x},t-\Delta t(x,\tilde{x},t))|^2\frac{\Delta \hat{\mathbf{r}}_1(x,\tilde{x},t)}{|\Delta \hat{\mathbf{r}}_1(x,\tilde{x},t)|^3}\sqrt{g(\mathbf{r}(\tilde{x},t-\Delta t(x,\tilde{x},t)))}d\tilde{x}.\nonumber\end{eqnarray}

Now, we define

\begin{equation}\label{br100} dF_{E'}=\left(
\begin{array}{llll}
 0 &  0 & 0 & -i_mdE_1'
 \\
 0& 0 & 0 & -i_mdE_2'
 \\
 0 & 0 & 0 & -i_mdE_3'
 \\
 i_m dE_1' & i_m dE_2' & i_m dE_3'& 0  \end{array} \right) \end{equation}
  and define $dF_E$ through the relations
 $$(dF_{E'})_{jk}=\hat{a}_{jl}\hat{a}_{kp}(dF_{E})_{lp},$$
 where
 $\{\hat{a}_{jk}\}$ is such that
 $$x_j'=\hat{a}_{jk}x_k,$$
 or more specifically,
 $$x_j'=x_j+\left(\frac{1}{\sqrt{1-\frac{v^2}{c^2}}}-1\right)\frac{\mathbf{x}\cdot \mathbf{v}}{v^2}v_j-\frac{tv_j}{\sqrt{1-\frac{v^2}{c^2}}},$$
$\forall j \in \{1,2,3\},$
and
$$x_4'/(i_mc)=t'=\frac{1}{\sqrt{1-\frac{v^2}{c^2}}}\left(t-\frac{\mathbf{x}\cdot \mathbf{v}}{c^2}\right),$$
where here,
$$\mathbf{v}=\frac{\partial \mathbf{r}_1(x,t)}{\partial t}.$$

At this point we assume a functional  $W(R(\mathbf{r}),\mathbf{r})$, which corresponds to the self-interacting energy, is such that
$$\delta_R W(R(\mathbf{r}),\mathbf{r})=|R(\mathbf{r})|\int_\Omega K \frac{dq(\tilde{x},t-\Delta t(x,\tilde{x},t))}{|\Delta \hat{\mathbf{r}}_1(x,\tilde{x},t)|}$$
and,
$$\delta_{\mathbf{r}} W(R(\mathbf{r}),\mathbf{r})=\delta_R W(R(\mathbf{r}),\mathbf{r}) \frac{\partial R(\mathbf{r})}
{\partial \mathbf{r}}+ |R(\mathbf{r}(x,t))|^2 \int_\Omega \sum_{k=1}^3 d \tilde{E}_k (x,\tilde{x},t) e_k$$
where $\{e_1,e_2,e_3\}$ is the canonical basis of $\mathbb{R}^3$ and we have denoted
\begin{equation}\label{br100} dF_{E}(x,\tilde{x},t)=\left(
\begin{array}{cccc}
 0 &  d\tilde{B}_3 & -d\tilde{B}_2 & -i_md\tilde{E}_1
 \\
 -d\tilde{B}_3& 0 & d\tilde{B}_1 & -i_md\tilde{E}_2
 \\
 d\tilde{B}_2 & -d\tilde{B}_1 & 0 & -i_m d\tilde{E}_3
 \\
 i_md\tilde{E}_1 & i_m d\tilde{E}_2 & i_m d\tilde{E}_3& 0,  \end{array} \right) \end{equation}

\begin{equation}\label{br101} F_E(x,t)=\int_\Omega dF_E(x,\tilde{x},t)=\left(
\begin{array}{cccc}
 0 &  \tilde{B}_3 & -\tilde{B}_2 & -i_m\tilde{E}_1
 \\
 -\tilde{B}_3& 0 & \tilde{B}_1 & -i_m\tilde{E}_2
 \\
 \tilde{B}_2 & -\tilde{B}_1 & 0 & -i_m\tilde{E}_3
 \\
 i_m\tilde{E}_1 & i_m \tilde{E}_2 & i_m\tilde{E}_3& 0  \end{array} \right) \end{equation}

where these last integrations are in $\tilde{x}$.

Also,
$$\tilde{\mathbf{E}}=(\tilde{E}_1,\tilde{E}_2,\tilde{E}_3),$$
and
$$\tilde{\mathbf{B}}=(\tilde{B}_1,\tilde{B}_2,\tilde{B}_3).$$

At this point, up to a concerning Lorentz transformation, we assume to be possible to express the total electric field $\mathbf{E}$ by
$$\mathbf{E}=\mathbf{\tilde{E}}+\nabla \Phi+\frac{1}{c}\frac{\partial \mathbf{A}}{\partial t},$$
for appropriate functions $\Phi \in W^{1,2}(\Omega \times [0,T])$ and $\mathbf{A} \in W^{1,2}(\Omega \times [0,T];\mathbb{R}^3).$

Also $$\mathbf{B}=\mathbf{B}_0+\tilde{\mathbf{B}}-\text{curl } \mathbf{A},$$ where $\mathbf{A}=(A_1,A_2,A_3)$ is a magnetic potential.

From the standard literature, we also define $A_4=i_m \;\Phi$
and assume the Lorentz condition,
$$\text{ div }\mathbf{A}+\frac{1}{c}\frac{\partial \Phi}{\partial t}=0, \text{ in } \Omega \times [0,T].$$

So, the system energy may be written as

\begin{eqnarray}
&&J(R,\mathbf{r},\mathbf{n},\Phi,\mathbf{A},E,\lambda)\nonumber \\ &=& c^2\int_0^T\int_\Omega |R(\mathbf{r})|^2\sqrt{1-\frac{v^2}{c^2}}\sqrt{-g}\;dx\;dt
\nonumber \\ && +\frac{1}{c}\int_0^T\int_\Omega K_1 |R(\mathbf{r})|^2 \frac{\partial \mathbf{r}_1}{\partial t} \cdot \mathbf{A} \;\sqrt{-g}\;dx\;dt
\nonumber \\ &&+\frac{\gamma}{2}\int_0^T\int_\Omega \hat{R} \sqrt{-g}\;dx\;dt
  \nonumber \\ &&+ W(R(\mathbf{r}),\mathbf{r}) \nonumber \\ &&+K_1\int_0^T\int_\Omega \Phi(\mathbf{r})  |R(\mathbf{r})|^2\;\sqrt{-g} dx\;dt \nonumber \\ &&+\frac{1}{8\pi}\left(\|\mathbf{B}\|_{\Omega \times [0,T],2}^2
  +\|\mathbf{E}\|_{\Omega \times [0,T],2}^2\right)\nonumber \\ &&-\frac{1}{2}\int_0^TE(t) \left(\int_\Omega |R(\mathbf{r})|^2\;\sqrt{-g}\;dx-m_e\right)\;dt
\nonumber \\ &&+\left\langle \lambda_1, \mathbf{n} \cdot \frac{\partial \mathbf{r}}{\partial \overline{t}} \right\rangle_{L^2}
+\left\langle \lambda_2, \mathbf{n} \cdot \mathbf{n}-1 \right\rangle_{L^2}\nonumber \\ &&+\left\langle \lambda_3,\text{div }\mathbf{A}+\frac{1}{c}\frac{\partial \Phi}{\partial t} \right\rangle_{L^2} .
\end{eqnarray}
Also,
$$\mathbf{g}_k=\frac{\partial\mathbf{r}(x,t)}{\partial \hat{x}_k},$$
where here $\{\hat{x}_k\}=(x_1,x_2,x_3,ct)$.
Moreover,
$$g=det\{g_{ij}\},$$ $$g_{ij}=\mathbf{g}_i \cdot \mathbf{g}_j,$$
where here again, such a  product is given by $$ \mathbf{y} \cdot \mathbf{z} =-y_4z_4+\sum_{i=1}^3y_i z_i, \; \forall \mathbf{y}=(y_1,y_2,y_3,y_4), \; \mathbf{z}=(z_1,z_2,z_3,z_4) \in \mathbb{R}^4,$$
$$\hat{R}=g^{ij}\hat{R}_{ij},$$
$$\hat{R}_{jk}=\hat{R}_{jik}^i,$$
$$\hat{R}_{jkl}^i=b^l_i\;b_{jk}^*,$$
$$b_{ij}=-\frac{1}{\sqrt{m}}\left(\frac{\partial \left(R(\mathbf{r})\mathbf{n}(\mathbf{r})\right)}{\partial \hat{x}_j}-i_m\;A_j\;R(\mathbf{r})  \mathbf{n}(\mathbf{r})\right)\cdot \mathbf{g}_i.$$
Furthermore,
$$b_j^i=g^{il}b_{lj},$$
and,
$$\{g^{ij}\}=\{g_{ij}\}^{-1},$$
$\forall i,j,k \in \{1,2,3,4\}.$

Finally, here we would also have $$v=\sqrt{\left(\frac{\partial X_1}{\partial t}\right)^2+\left(\frac{\partial X_2}{\partial t}\right)^2+\left(\frac{\partial X_3}{\partial t}\right)^2},$$
Its worth mentioning $\lambda_1,\lambda_2,\lambda_3$ are appropriate Lagrange multipliers concerning the respective constraints.
\begin{remark} It is possible in some cases we cannot find
$W(R(\mathbf{r}),\mathbf{r})$, which corresponds to the self-interacting energy, such that
$$\delta_R W(R(\mathbf{r}),\mathbf{r})=|R(\mathbf{r})|\int_\Omega K \frac{dq(\tilde{x},t-\Delta t(x,\tilde{x},t))}{|\Delta \hat{\mathbf{r}}_1(x,\tilde{x},t)|}$$
and,
$$\delta_{\mathbf{r}} W(R(\mathbf{r}),\mathbf{r})=\delta_R W(R(\mathbf{r}),\mathbf{r}) \frac{\partial R(\mathbf{r})}
{\partial \mathbf{r}}+ |R(\mathbf{r}(x,t))|^2 \int_\Omega\sum_{k=1}^3 d \tilde{E}_k (x,\tilde{x},t)e_k.$$

In such a case, such last equations may be just approximately satisfied, so that we could define an optimization problem corresponding to find a critical point of the functional $J$ plus a positive constant multiplied by the $L^2$ norms of analytical expressions corresponding to these last two equations.
\end{remark}

\section{A new  interpretation of the Bohr atomic model}

This section develops a new  interpretation of Bohr atomic model through classical and quantum mechanics.

In a second step, we consider as a generalization of such a model, the issue of an interacting system comprised by a large amount of
same type atoms.

At this point we start to describe such a model.

Let $\Omega =B_{R_0}(\mathbf{0}) \subset \mathbb{R}^3$ be an open ball with center at $\mathbf{0} \in \mathbb{R}^3.$ Let $[0,T]$
be a time interval. For $n \in \mathbb{N}$, consider a system with $\sum_{l=0}^{n-1}(2l+1)$ electrons and the same number of protons,
where the protons are supposed to be at rest at $\mathbf{x}=\mathbf{0} \in \mathbb{R}^3$. Moreover, the electrons are distributed in $n$ layers
$l \in \{0, \ldots, n-1\},$ each layer $l$ with $2l+1$ electrons.

We denote the position field for the electron $j\in -l,\ldots,0, \ldots, l$ at the layer $l$, by
$\mathbf{r}_j^l:\Omega \times [0,T] \rightarrow \mathbb{R}^3$, where
\begin{eqnarray}\label{eq128}\mathbf{r}_j^l(\mathbf{x},t)&=&R_j^l(r)\left(\sin( (w_1)_j^l(\mathbf{x}) t +\theta_j^l)\cos( (w_2)_j^l(\mathbf{x}) t +\phi_j^l)\mathbf{i}
\right.\nonumber \\ &&\left.+\sin( (w_1)_j^l(\mathbf{x}) t +\theta_j^l)\sin( (w_2)_j^l(\mathbf{x}) t +\phi_j^l)\mathbf{j}
 +\cos( (w_1)_j^l(\mathbf{x}) t +\theta_j^l) \mathbf{k}\right).\end{eqnarray}

We also recall that $ \mathbf{x} \in \mathbb{R}^3$ in spherical coordinates corresponds to $(r,\theta,\phi)$
and $\{\mathbf{i},\mathbf{j},\mathbf{k}\}$ is the canonical basis of $\mathbb{R}^3.$

Moreover, the density scalar field for such a same electron is denoted by $ m_e |\varphi_j^l|^2:\Omega \rightarrow \mathbb{R}$, where
$$\varphi_j^l(\mathbf{x})=\hat{\varphi}_j^l(r) (L^-)^{(l+j)}[(\sin \theta)^le^{i\phi l}],$$
 $i$ denotes the imaginary unit,
$$L_x=-\frac{\hbar}{i}\left(\sin \phi \frac{\partial }{\partial \theta}+\cot \theta \cos \phi \frac{\partial}{\partial \phi}\right),$$
$$L_y=\frac{\hbar}{i}\left(\cos \phi \frac{\partial }{\partial \theta}-\cot \theta \sin \phi \frac{\partial}{\partial \phi}\right)$$
and
$$L^{-}=\frac{1}{\hbar}(L_x-iL_y),$$ and where $(L^-)^0=I_d$ (identity operator).
\begin{remark}

In principle, we would expect $\mathbf{r}_j^l$ to be an injective function, so that $$\tilde{\varphi}_j^l(\mathbf{r}_j^l(\mathbf{x},t))=
\varphi_j^l(\mathbf{x},t)=\varphi_j^l((\mathbf{r}_j^l)^{-1}(\mathbf{r}_j^l(\mathbf{x},t)))$$ is well defined.

This may not be the case for the motion indicated in (\ref{eq128}).
Thus, such a concerning  motion suggests us a new interpretation of the Bohr atomic model and related wave particle duality
for the electrons in the atom in question.\end{remark}

We also define,
$$U=\{\varphi=\{\varphi_j^l\} \in W^{1,2}(\Omega;\mathbb{C}^{Z})\;:\; \varphi_j^l=0\text{ on } \partial \Omega\},$$
and
 $$V=\{\mathbf{r}=\{\mathbf{r}_j^l\} \in W^{1,2}(\Omega \times [0,T]; \mathbb{R}^{3Z})\;:\;R_j^l(0)=0, \text{ and } R_j^l(R_0)=R_0\}.$$

For such a system, we consider the following types of energy.

\begin{enumerate}

\item Kinetics energy, denoted by $E_c$, where
$$E_c=\frac{1}{2}\sum_{l=0}^{n-1} \sum_{j=-l}^l \int_0^T\int_\Omega m_e |\varphi_j^l(\mathbf{x})|^2\frac{\partial \mathbf{r}_j^l(\mathbf{x},t)}{\partial t}
\cdot  \frac{\partial \mathbf{r}_j^l(\mathbf{x},t)}{\partial t} \;d\mathbf{x} dt,$$
where $m_e$ denotes the mass of a single electron and $$d\mathbf{x}=dx_1dx_2dx_3.$$

\item A regularizing part for the position field, denoted by $E_r$, where
$$E_r=\frac{1}{2}\sum_{l=0}^{n-1} \sum_{j=-l}^l\sum_{k=1}^3\int_0^T \int_\Omega A^l |\varphi_j^l(\mathbf{x})|^2\frac{\partial \mathbf{r}_j^l(\mathbf{x},t)}{\partial x_k}
\cdot  \frac{\partial \mathbf{r}_j^l(\mathbf{x},t)}{\partial x_k} \;d\mathbf{x} dt,$$
with $A^l>0$ to be specified, $\forall l \in \{0,\ldots,n-1\}.$

\item Coulomb electronic interaction (classical), denoted by $E_{int}$, where
in a first approximation, we consider only the interaction for the same layer electrons, neglecting the interactions between
different layer electrons.

Thus,
$$E_{int}=\frac{1}{4}\sum_{l=0}^{n-1} \sum_{j=-l}^l \sum_{k=-l}^l Ke^2
\int_0^T\int_\Omega\int_\Omega\frac{ |\varphi_j^l(\mathbf{x})|^2|\varphi_k^l(\tilde{\mathbf{x}})|^2}{|\mathbf{r}_j^l(\mathbf{x},t)-\mathbf{r}_k^l(\tilde{\mathbf{x}},t)|}\;d\mathbf{x} d\tilde{\mathbf{x}}dt,$$
where $e$ is the charge of a single electron and $K>0$ is a an appropriate constant to be specified.

\item Coulomb interaction of each electron with the heavier nucleus, denoted by $E_{int}^{p}$,
where
$$E_{in}^{p}=\frac{1}{2}\sum_{l=0}^{n-1} \sum_{j=-l}^l K e^2 Z\int_0^T\int_\Omega  \frac{|\varphi_j^l(\mathbf{x})|^2}{|\mathbf{r}_j^l(\mathbf{x},t)|}\;d\mathbf{x}dt.$$

\item Energy related to the presence of external potentials $V_j^l$, denoted by $E_p$, where
$$E_p= \frac{1}{2}\sum_{l=0}^{n-1}\sum_{j=-l}^l\int_0^T\int_\Omega V_j^l(\mathbf{x})|\varphi_j^l(\mathbf{x})|^2\;d\mathbf{x}dt.$$

\item A regularizing and curvature distribution control term for the scalar density field (quantum part), denoted by $E_q$, where
$$E_q=\frac{1}{2}\sum_{l=0}^{n-1} \sum_{j=-l}^l \sum_{k=1}^3
\gamma_j^l \int_0^T\int_\Omega \frac{\partial (\varphi_j^l(\mathbf{x})\mathbf{n}_j^l(\mathbf{x},t))}{\partial x_k}\cdot \frac{\partial (\varphi_j^l(\mathbf{x})\mathbf{n}_j^l(\mathbf{x},t))}{\partial x_k}\;d\mathbf{x}dt,$$
where the normal field $\mathbf{n}^l_j$ may be given by
\begin{eqnarray}\mathbf{n}_j^l(\mathbf{x},t)&=&\sin( (w_1)_j^l(\mathbf{x}) t +\theta_j^l)\cos( (w_2)_j^l(\mathbf{x}) t +\phi_j^l)\mathbf{i}
\nonumber \\ &&+\sin( (w_1)_j^l(\mathbf{x}) t +\theta_j^l)\sin( (w_2)_j^l(\mathbf{x}) t +\phi_j^l)\mathbf{j}
 +\cos( (w_1)_j^l(\mathbf{x}) t +\theta_j^l) \mathbf{k},\end{eqnarray}
 so that,
$$\mathbf{n}_j^l(\mathbf{x},t) \cdot \frac{\partial \mathbf{r}_j^l(\mathbf{x},t)}{\partial t}=0, \text{ in } \Omega \times [0,T],$$
$\forall j \in \{-l, \ldots,0, \ldots,l\},\;\forall l \in \{0, \ldots,n-1\}.$

\item Constraints: The system is subject to the following constraints,
$$\int_\Omega |\varphi_j^l(\mathbf{x})|^2\;dx=1,\;\forall l \in \{0,\ldots,n-1\},\; j \in \{-l,\ldots,0,\ldots, l\}.$$
\end{enumerate}
%\item $$\mathbf{n}_j^l(\mathbf{x},t) \cdot \frac{\partial \mathbf{r}_j^l(\mathbf{x},t)}{\partial t}=0, \text{ in } \Omega \times [0,T],$$
%$\forall l \in \{0,\ldots,n-1\},\; j \in \{-l,\ldots,0,\ldots, l\}.$

Hence, the total system energy is given by the functional $J:U\times V \times \mathbb{R}^Z \rightarrow \mathbb{R}$ where already including the Lagrange multipliers,
we have
\begin{eqnarray}
J(\varphi,\mathbf{r},E)&=& -E_c+E_r+E_{in}-E_{in}^p+E_p+E_q \nonumber \\ &&-\frac{1}{2}\sum_{l=0}^{n-1} \sum_{j=-l}^l
E_j^lT \left(\int_\Omega |\varphi_j^l(\mathbf{x})|^2\;d\mathbf{x}-1\right).
\end{eqnarray}
Summarizing,
\begin{eqnarray}
&&J(\varphi,\mathbf{r},E)\nonumber \\ &=& -\frac{1}{2}\sum_{l=0}^{n-1} \sum_{j=-l}^l \int_0^T\int_\Omega m_e |\varphi_j^l(\mathbf{x})|^2\frac{\partial \mathbf{r}_j^l(\mathbf{x},t)}{\partial t}
\cdot  \frac{\partial \mathbf{r}_j^l(\mathbf{x},t)}{\partial t} \;d \mathbf{x} dt \nonumber \\ &&+\frac{1}{2}
\sum_{l=0}^{n-1} \sum_{j=-l}^l\sum_{k=1}^3\int_0^T \int_\Omega A^l |\varphi_j^l(\mathbf{x})|^2\frac{\partial \mathbf{r}_j^l(\mathbf{x},t)}{\partial x_k}
\cdot  \frac{\partial \mathbf{r}_j^l(\mathbf{x},t)}{\partial x_k} \;d\mathbf{x} dt \nonumber \\ &&+
\frac{1}{4}\sum_{l=0}^{n-1} \sum_{j=-l}^l \sum_{k=-l}^l Ke^2
\int_0^T\int_\Omega\int_\Omega\frac{ |\varphi_j^l(\mathbf{x})|^2|\varphi_k^l(\tilde{\mathbf{x}})|^2}{|\mathbf{r}_j^l(\mathbf{x},t)-\mathbf{r}_k^l(\tilde{\mathbf{x}},t)|}\;d \mathbf{x}d \mathbf{\tilde{x}}dt
\nonumber \\ &&-
\frac{1}{2}\sum_{l=0}^{n-1} \sum_{j=-l}^l K e^2 Z\int_0^T\int_\Omega  \frac{|\varphi_j^l(\mathbf{x})|^2}{|\mathbf{r}_j^l(\mathbf{x},t)|}\;d\mathbf{x}dt
\nonumber \\ &&+
\frac{1}{2}\sum_{l=0}^{n-1} \sum_{j=-l}^l \sum_{k=1}^3
\gamma_j^l \int_0^T\int_\Omega \frac{\partial (\varphi_j^l(\mathbf{x})\mathbf{n}_j^l(\mathbf{x},t))}{\partial x_k}\cdot \frac{\partial (\varphi_j^l(\mathbf{x})\mathbf{n}_j^l(\mathbf{x},t))}{\partial x_k}\;d\mathbf{x}dt
\nonumber \\ &&-\frac{1}{2}\sum_{l=0}^{n-1} \sum_{j=-l}^l
E_j^lT \left(\int_\Omega |\varphi_j^l(\mathbf{x})|^2\;d\mathbf{x}-1\right).
\end{eqnarray}
With such statements and definitions in mind, we define the control problem of finding $\{\theta_j^l,\phi_j^l\} \in \mathbb{R}^Z \times \mathbb{R}^Z,$
which minimizes
$$J_1(\varphi,\mathbf{r},E\})$$
where
\begin{eqnarray}
&&J_1(\varphi,\mathbf{r},E\})\nonumber \\ &=&
\frac{1}{4}\sum_{l=0}^{n-1} \sum_{j=-l}^l \sum_{k=-l}^l Ke^2
\int_0^T\int_\Omega\int_\Omega\frac{ |\varphi_j^l(\mathbf{x})|^2|\varphi_k^l(\tilde{\mathbf{x}})|^2}{|\mathbf{r}_j^l(\mathbf{x},t)-\mathbf{r}_k^l(\tilde{\mathbf{x}},t)|}\; d \mathbf{x} d \mathbf{\tilde{x}}dt,
\end{eqnarray}
subject to
\begin{enumerate}
\item
\begin{eqnarray}&& m_e|\varphi_j^l(\mathbf{x})|^2\frac{\partial^2 \mathbf{r}_j^l(\mathbf{x},t)}{\partial t^2}
\nonumber \\ &&-A^l \sum_{k=1}^3\frac{\partial}{\partial x_k}\left( |\varphi_j^l(\mathbf{x})|^2\frac{\partial \mathbf{r}_j^l(\mathbf{x},t))}{\partial x_k}\right) \nonumber \\ &&-\sum_{k=-l}^l|\varphi_j^l(\mathbf{x})|^2 \int_\Omega \frac{Ke^2|\varphi_k^l(\tilde{\mathbf{x}})|^2 (\mathbf{r}_j^l(\mathbf{x},t)-\mathbf{r}_k^l(\tilde{\mathbf{x}},t))}{|\mathbf{r}_j^l(\mathbf{x},t)-\mathbf{r}_k^l(\tilde{\mathbf{x}},t)|^3}\;d\tilde{\mathbf{x}}
\nonumber \\ &&+KZe^2  \frac{\varphi_j^l(\mathbf{x})|^2}{|\mathbf{r}_j^l(\mathbf{x},t)|^3}\;\mathbf{r}_j^l(\mathbf{x},t) \nonumber \\ &=& \mathbf{0}, \text{ in }
\Omega \times [0,T],
\end{eqnarray}

\item
\begin{eqnarray}
&& -m_e \hat{\varphi}_j^l(r)\frac{1}{T} \int_0^T \frac{\partial \mathbf{r}_j^l(\mathbf{x},t)}{\partial t} \cdot \frac{\partial \mathbf{r}_j^l(\mathbf{x},t)}{\partial t} \;dt  + \nonumber \\ &&
-\gamma_j^l \frac{1}{r^2} \frac{\partial }{\partial r}\left( r^2\frac{\partial \hat{\varphi}_j^l(r)}{\partial r}\right)
\nonumber \\ &&+ \gamma_j^l \frac{l(l+1)}{r^2} \hat{\varphi}_j^l(r) \nonumber \\ &&+\gamma_j^l \hat{\varphi}_j^l(r)
\frac{1}{T} \int_0^T \sum_{k=1}^3\left(\frac{\partial \mathbf{n}_j^l}{\partial x_k} \cdot \frac{\partial \mathbf{n}_j^l}{\partial x_k}\right)\;dt
\nonumber \\ &&
+\sum_{k=-l}^l\hat{\varphi}_j^l(r) \frac{1}{T}\int_0^T\int_\Omega\frac{Ke^2|\varphi_k^l(\tilde{\mathbf{x}})|^2}{|\mathbf{r}_j^l(\mathbf{x},t)-\mathbf{r}_k^l(\tilde{\mathbf{x}},t)|}
\;d\tilde{\mathbf{x}}dt
\nonumber \\ &&-KZe^2 \frac{\hat{\varphi}_j^l(r)}{R_j^l(r)} \nonumber \\ &&
-E_j^l\hat{\varphi}_j^l(r)= 0, \text{ in } [0,R_0],
\end{eqnarray}
and up to a normalizing constant for  $\varphi(\mathbf{x})$,
\item $$\int_0^{R_0} |\hat{\varphi}_j^l(r)|^2r^2\;dr=1,
$$
$\forall j \in \{-l, \ldots,0, \ldots,l\},\;\forall l \in \{0, \ldots,n-1\}.$
\end{enumerate}

\section{ A system with a large number of interacting atoms}

Now consider a system with a large number $N$ of interacting same type atoms, each one with $Z=\sum_{l=0}^{n-1} (2l+1)$ electrons and the same number of
protons.

Consider also the problem of finding  the $N$ nucleus positions, each one comprised by $Z$ protons, in an open,
bounded, connected set $\Omega \subset \mathbb{R}^3$ with a Lipschitzian boundary denoted by $\partial \Omega.$

We define the position field for the electron $j$, in the layer $l$ at the atom $k$, which the nucleus is located at $\mathbf{x}_k \in \Omega,$
denoted by $\mathbf{r}_j^l(\cdot,\mathbf{x}_k,\cdot):\Omega \times [0,T] \rightarrow \mathbb{R}^3$, as
\begin{eqnarray}
\mathbf{r}_j^l(\mathbf{x},\mathbf{x}_k,t)&=& \mathbf{x}_k
+\mathbf{R}_j^l(\mathbf{x},\mathbf{x}_k)e^{iw_j^l(\mathbf{x},\mathbf{x}_k)t}
\end{eqnarray}

Also, the respective density scalar field is denoted by $\varphi=\{\varphi_j^l:\Omega \rightarrow \mathbb{C}\}$,

Here $$\varphi(\cdot,\mathbf{x}_k) \in U, \forall k \in \{1,\ldots,N\}$$ and $$\mathbf{r}(\cdot,\mathbf{x}_k,\cdot) \in V,\; \forall k \in \{1,\ldots, N\}.$$

With such statements in mind, we consider the control problem of finding $\{\mathbf{x}_k\}_{k=1}^N$ and $ \{w_j^l(\mathbf{x},\mathbf{x}_k)\}$
 which minimizes $J_2+J_3+J_4$, where
\begin{eqnarray}
J_2&=& \sum_{l_1=0}^{n-1} \sum_{l_2=0}^{n-1} \sum_{j=-l_1}^{l_1} \sum_{k=-l_2}^{l_2} \sum_{k_1=1}^N \sum_{k_2=1}^N
Ke^2 \times \nonumber \\ && \times \int_0^T\int_\Omega\int_\Omega \frac{|\varphi_j^{l_1}(\mathbf{x},\mathbf{x}_{k_1})|^2|\varphi_k^{l_2}(\tilde{\mathbf{x}},\mathbf{x}_{k_2})|^2}
{|\mathbf{r}_j^{l_1}(\mathbf{x},\mathbf{x}_{k_1},t)-\mathbf{r}_k^{l_2}(\tilde{\mathbf{x}},\mathbf{x}_{k_2},t)|}\;d\mathbf{x} d\tilde{\mathbf{x}}dt
\end{eqnarray}
\begin{eqnarray}
J_3&=& \sum_{l=0}^{n-1}\sum_{j=-l}^l\sum_{k=1}^N \sum_{k_1=1}^{N}Ke^2Z\int_0^T\int_\Omega \frac{|\varphi_j^{l}(\mathbf{x},\mathbf{x}_k)|^2}{|\mathbf{r}_j^l(\mathbf{x},\mathbf{x}_k,t)-\mathbf{x}_{k_1}|}\;d\mathbf{x}dt,
\end{eqnarray}
and
$$J_4=\sum_{k=1}^N\sum_{k_1=1}^N \frac{Ke^2 Z^2}{|\mathbf{x}_k-\mathbf{x}_{k_1}|},$$

subject to
\begin{enumerate}
\item
\begin{eqnarray}
&&m_e |\varphi_j^l(\mathbf{x},\mathbf{x}_k)|^2\frac{\partial^2 \mathbf{r}_j^l(\mathbf{x},\mathbf{x}_k,t)}{\partial t^2}
\nonumber \\ && -\sum_{s=1}^3 A^l \frac{\partial }{\partial x_s} \left( |\varphi(\mathbf{x},\mathbf{x}_k)|^2 \frac{\partial \mathbf{r}_j^l(\mathbf{x},\mathbf{x}_k,t)}{\partial x_s}\right) \nonumber \\ &&-
\sum_{l_1=0}^{n-1} \sum_{p=-l_1}^{l_1} \sum_{k_1=1}^N Ke^2|\varphi_j^l(\mathbf{x},\mathbf{x}_k)|^2\int_\Omega
\frac{|\varphi_p^{l_1}(\tilde{\mathbf{x}},\mathbf{x}_{k_1})|^2(\mathbf{r}_j^l(\mathbf{x},\mathbf{x}_k,t)
-\mathbf{r}_p^{l_1}(\tilde{\mathbf{x}},\mathbf{x}_{k_1},t))}{|\mathbf{r}_j^l(\mathbf{x},\mathbf{x}_k,t)-\mathbf{r}_p^{l_1}(\tilde{\mathbf{x}},\mathbf{x}_{k_1},t)|^3}
\;d\tilde{\mathbf{x}} \nonumber \\ && +K e^2 Z\sum_{k_1=1}^N \frac{|\varphi_j^l(\mathbf{x},\mathbf{x}_k)|^2(\mathbf{r}_j^l(\mathbf{x},\mathbf{x}_k,t)-\mathbf{x}_{k_1})}
{|\mathbf{r}_j^l(\mathbf{x},\mathbf{x}_k,t)-\mathbf{x}_{k_1}|^3}
\nonumber \\ &=& \mathbf{0}, \text{ in } \Omega.
\end{eqnarray}
\item
$$\mathbf{n}_j^l(\mathbf{x},\mathbf{x}_k,t) \cdot \frac{\partial\mathbf{r}_j^l(\mathbf{x},\mathbf{x}_k,t)}{\partial t}=0, \text{ in } \Omega \times [0,T],$$
\item   $$\mathbf{n}_j^l(\mathbf{x},\mathbf{x}_k,t) \cdot \mathbf{n}_j^l(\mathbf{x},\mathbf{x}_k,t)=1, \text{ in } \Omega \times [0,T],$$
\item
\begin{eqnarray}
&&-\sum_{s=1}^3 \gamma_j^l \frac{\partial^2 \varphi_j^l(\mathbf{x},\mathbf{x}_k)}{\partial x_s^2}
\nonumber \\ && +\gamma_j^l {\varphi}_j^l(\mathbf{x},\mathbf{x}_k)
\frac{1}{T} \int_0^T \sum_{s=1}^3\left(\frac{\partial \mathbf{n}_j^l(\mathbf{x},\mathbf{x}_k,t)}{\partial x_s} \cdot \frac{\partial \mathbf{n}_j^l(\mathbf{x},\mathbf{x}_k,t)}{\partial x_s}\right)\;dt \nonumber \\ &&+\varphi_j^l(\mathbf{x},\mathbf{x}_k) \sum_{l_1=0}^{n-1} \sum_{k=-l_1}^{l_1} \sum_{k_1=1}^N\frac{1}{T} \int_\Omega
\frac{|\varphi_k^{l_1}(\tilde{\mathbf{x}},\mathbf{x}_{k_1})|^2}
{|\mathbf{r}_j^l(\mathbf{x},\mathbf{x}_k)-\mathbf{r}_k^{l_1}(\tilde{\mathbf{x}},\mathbf{x}_{k_1},t)|}\;d\tilde{\mathbf{x}}dt
\nonumber \\ && -\sum_{k_1=1}^N Ke^2Z \frac{\varphi_j^l(\mathbf{x},\mathbf{x}_k)}{|\mathbf{r}(\mathbf{x},\mathbf{x}_k,t)-\mathbf{x}_{k_1}|}
\nonumber \\ && +\sum_{s=1}^3A^l \varphi_j^l(\mathbf{x},\mathbf{x}_k)\frac{1}{T} \int_0^T\frac{\partial \mathbf{r}_j^l(\mathbf{x},\mathbf{x}_k,t)}{\partial x_s}\cdot \frac{\partial \mathbf{r}_j^l(\mathbf{x},\mathbf{x}_k,t)}{\partial x_s}\;dt
\nonumber \\  && -m_e \varphi_j^l(\mathbf{x},\mathbf{x}_k)\frac{1}{T} \int_0^T\frac{\partial \mathbf{r}_j^l(\mathbf{x},\mathbf{x}_k,t)}{\partial t}\cdot \frac{\partial \mathbf{r}_j^l(\mathbf{x},\mathbf{x}_k,t)}{\partial t} \;dt\nonumber \\ &&-E_{jk}^l \varphi_j^l(\mathbf{x},\mathbf{x}_k)=0, \text{ in } \Omega.
\end{eqnarray}
\item
$$\int_\Omega |\varphi_j^l(\mathbf{x},\mathbf{x}_k)|^2\;dx=1,\; \forall l \in \{0,\ldots,n-1\},\; j \in \{-l,\ldots,0,\ldots,l\},\; k \in \{1,\dots,N\}.$$
\end{enumerate}
\subsection{ A proposal for the case in which $N$ is very large}

As $N$ is very large, we shall propose a limit density scalar field $\varphi_j^l(\mathbf{x}, \mathbf{y})$, that is
$$\varphi_j^l:\Omega \times \Omega \rightarrow \mathbb{C}.$$

Also, we shall propose, as the position vector field, $\mathbf{r}_j^l(\mathbf{x},\mathbf{y},t)$, that is
$\mathbf{r}_j^l: \Omega \times \Omega \times [0,T]\rightarrow \mathbb{R}^3$, where
%\begin{eqnarray}
%\mathbf{r}_j^l(\mathbf{x},\mathbf{y},t)&=& \mathbf{y}
%\nonumber \\ &&+R_j^l(|\mathbf{x}-\mathbf{y}|)
%\nonumber \\ &&(\sin(\theta^l_{j} +\hat{\theta}^l(\mathbf{y}))\cos(\phi_j^l+\hat{\phi}^l(\mathbf{y}))\mathbf{i}
%\nonumber \\ &&+ \sin(\theta^l_{j} +\hat{\theta}^l(\mathbf{y}))\sin(\phi_j^l+\hat{\phi}^l(\mathbf{y}))\mathbf{j}
%\nonumber \\ &&+\cos(\theta_j^l+\hat{\theta}^l(\mathbf{y}))\mathbf{k})
%\nonumber \\ &&+\mathbf{R}(\mathbf{x},\mathbf{y})e^{iw(\mathbf{x},\mathbf{y})t}
%\end{eqnarray}
\begin{eqnarray}
\mathbf{r}_j^l(\mathbf{x},\mathbf{y},t)&=& \mathbf{y}
%\nonumber \\ &&+R_j^l(|\mathbf{x}-\mathbf{y}|)
%\nonumber \\ &&(\sin(\theta^l_{j} +\hat{\theta}^l(\mathbf{y}))\cos(\phi_j^l+\hat{\phi}^l(\mathbf{y}))\mathbf{i}
%\nonumber \\ &&+ \sin(\theta^l_{j} +\hat{\theta}^l(\mathbf{y}))\sin(\phi_j^l+\hat{\phi}^l(\mathbf{y}))\mathbf{j}
%\nonumber \\ &&+\cos(\theta_j^l+\hat{\theta}^l(\mathbf{y}))\mathbf{k})
%\nonumber \\ &&
+\mathbf{R}_j^l(\mathbf{x},\mathbf{y})e^{iw_j^l(\mathbf{x},\mathbf{y})t}.
\end{eqnarray}

 We assume $\varphi=\{\varphi_j^l\} \in U$, where
$$U=\{\varphi=\{\varphi_j^l\} \in W^{1,2}(\Omega \times \Omega ; \mathbb{C}^Z)\;:\; \varphi_j^l=0 \text{ on } \partial  (\Omega \times  \Omega)\}$$ and $\mathbf{r}=\{\mathbf{r}_j^l\} \in V$, where here
$$V=\{\mathbf{r}=\{\mathbf{r}_j^l\} \in W^{1,2}(\Omega \times \Omega \times [0,T] ;\mathbb{R}^{3Z}) \;:\; \mathbf{R}_j^l=\mathbf{0} \text{ on } \partial (\Omega
\times \Omega)\}.$$

 For the protons, we specify the density scalar field $\varphi_p:\Omega \times \Omega
\rightarrow \mathbb{C},$ and the respective position field $\mathbf{r}_p(\mathbf{x},\mathbf{y})=\mathbf{y}$.
Moreover, $\varphi_p \in U_p$, where $$U_p=\{\varphi_p \in W^{1,2}(\Omega \times \Omega;\mathbb{C})\;:\; \varphi_p(\mathbf{x},\mathbf{y})=0, \text{ in } \partial (\Omega \times \Omega)\}.
$$

 In the distributional sense, we should approximately expect to obtain $$\varphi_p(\mathbf{x},\mathbf{y})=\delta(\mathbf{x}-\mathbf{y}), \text{ in } (\Omega \times \Omega)^0$$
where $(\Omega \times \Omega)^0$ denotes the interior of $\Omega\times \Omega$. Also, $\delta(\mathbf{x}-\mathbf{y})$ denotes a standard Dirac delta.

With such statements in mind, we consider the control problem of finding $\{w_j^l\}$ which minimizes $J_2+J_3+J_4$, where
\begin{eqnarray}
J_2&=& \sum_{l_1=0}^{n-1} \sum_{l_2=0}^{n-1} \sum_{j=-l_1}^{l_1} \sum_{k=-l_2}^{l_2} Ke^2 \times \nonumber \\ && \times \int_0^T\int_\Omega\int_\Omega
\int_\Omega \int_\Omega \frac{|\varphi_j^{l_1}(\mathbf{x},\mathbf{y})|^2|\varphi_k^{l_2}(\tilde{\mathbf{x}},\tilde{\mathbf{y}})|^2}
{|\mathbf{r}_{j}^{l_1}(\mathbf{x},\mathbf{y},t)-\mathbf{r}_k^{l_2}(\tilde{\mathbf{x}},\tilde{\mathbf{y}},t)|}\;d\mathbf{x} d\tilde{\mathbf{x}}d \mathbf{y} d\tilde{\mathbf{y}}dt
\end{eqnarray}
\begin{eqnarray}
J_3&=& \sum_{l=0}^{n-1}\sum_{j=-l}^l\sum_{k=1}^NKe^2Z\int_0^T\int_\Omega \int_\Omega \int_\Omega \frac{|\varphi_j^{l}(\mathbf{x},\mathbf{y})|^2}{|\mathbf{r}_j^l(\mathbf{x},\mathbf{y},t)-\tilde{\mathbf{y}}|}\;d\mathbf{x} d \mathbf{y}d\tilde{\mathbf{y}}dt,
\end{eqnarray}
and
$$J_4=T \int_\Omega \int_\Omega \int_\Omega \int_\Omega \frac{Ke^2 Z^2|\varphi_p(\mathbf{x},\mathbf{y})|^2|\varphi_p(\tilde{\mathbf{x}},\tilde{\mathbf{y}})|^2}{|\mathbf{y}-\tilde{\mathbf{y}}|}\;d\mathbf{x} d\tilde{\mathbf{x}}d\mathbf{y}d \tilde{\mathbf{y}},$$

subject to
\begin{enumerate}
\item
\begin{eqnarray}
&&m_e |\varphi_j^l(\mathbf{x},\mathbf{y})|^2\frac{\partial^2 \mathbf{r}_j^l(\mathbf{x},\mathbf{y},t)}{\partial t^2}
\nonumber \\ && -A^l\sum_{s=1}^3 \left( \frac{\partial }{\partial x_s} \left( |\varphi(\mathbf{x},\mathbf{y})|^2 \frac{\partial \mathbf{r}_j^l(\mathbf{x},\mathbf{y},t)}{\partial x_s}\right)\right.
\nonumber \\ &&+\left.\frac{\partial }{\partial y_s} \left( |\varphi(\mathbf{x},\mathbf{y})|^2\frac{\partial \mathbf{r}_j^l(\mathbf{x},\mathbf{y},t)}{\partial y_s}\right)\right) \nonumber \\ &&-
\sum_{l_1=0}^{n-1} \sum_{s=-l_1}^{l_1}  Ke^2|\varphi_j^l(\mathbf{x},\mathbf{y})|^2\int_\Omega\int_\Omega
\frac{|\varphi_s^{l_1}(\tilde{\mathbf{x}},\tilde{\mathbf{y}})|^2(\mathbf{r}_j^l(\mathbf{x},\mathbf{y},t)
-\mathbf{r}_s^{l_1}(\tilde{\mathbf{x}},\tilde{\mathbf{y}},t))}{|\mathbf{r}_j^l(\mathbf{x},\mathbf{y},t)-\mathbf{r}_s^{l_1}(\tilde{\mathbf{x}},\tilde{\mathbf{y}},t)|^3}
\;d\tilde{\mathbf{x}}d\tilde{\mathbf{y}} \nonumber \\ && +K e^2 Z|\varphi_j^l(\mathbf{x},\mathbf{y})|^2\int_\Omega \frac{(\mathbf{r}_j^l(\mathbf{x},\mathbf{y},t)-\tilde{\mathbf{y}})}
{|\mathbf{r}_j^l(\mathbf{x},\mathbf{y},t)-\tilde{\mathbf{y}}|^3} d\tilde{\mathbf{y}}
\nonumber \\ &=& \mathbf{0}, \text{ in } \Omega\times \Omega \times[0,T].
\end{eqnarray}
\item
$$\mathbf{n}_j^l(\mathbf{x},\mathbf{y},t) \cdot \frac{\partial \mathbf{r}_j^l(\mathbf{x},\mathbf{y},t)}{\partial t}=0, \text{ in } \Omega \times
\Omega \times  [0,T],$$
\item   $$\mathbf{n}_j^l(\mathbf{x},\mathbf{y},t) \cdot \mathbf{n}_j^l(\mathbf{x},\mathbf{y},t)=1, \text{ in } \Omega \times \Omega \times [0,T],$$
\item
\begin{eqnarray}
&&-\gamma_j^l \sum_{s=1}^3 \left(\frac{\partial^2 \varphi_j^l(\mathbf{x},\mathbf{y})}{\partial x_s^2}+  \frac{\partial^2 \varphi_j^l(\mathbf{x},\mathbf{y})}{\partial y_s^2}\right)
\nonumber \\ && +\gamma_j^l {\varphi}_j^l(\mathbf{x},\mathbf{y})
\frac{1}{T} \int_0^T \sum_{s=1}^3\left(\left(\frac{\partial \mathbf{n}_j^l(\mathbf{x},\mathbf{y},t)}{\partial x_s} \cdot \frac{\partial \mathbf{n}_j^l(\mathbf{x},\mathbf{y},t)}{\partial x_s}\right)+\left(\frac{\partial \mathbf{n}_j^l(\mathbf{x},\mathbf{y},t)}{\partial y_s} \cdot \frac{\partial \mathbf{n}_j^l(\mathbf{x},\mathbf{y},t)}{\partial y_s}\right)\right)\;dt
\nonumber \\ &&+\varphi_j^l(\mathbf{x},\mathbf{y}) \sum_{l_1=0}^{n-1} \sum_{k=-l_1}^{l_1} \frac{1}{T} \int_0^T \int_\Omega\int_\Omega
\frac{|\varphi_k^{l_1}(\tilde{\mathbf{x}},\tilde{\mathbf{y}})|^2}
{|\mathbf{r}_j^l(\mathbf{x},\mathbf{y},t)-\mathbf{r}_k^{l_1}(\tilde{\mathbf{x}},\tilde{\mathbf{y}},t)|}\;d\tilde{\mathbf{x}}d\tilde{\mathbf{y}}dt
\nonumber \\ && -\sum_{k_1=1}^N Ke^2Z \;\varphi_j^l(\mathbf{x},\mathbf{y})\int_\Omega\int_\Omega\frac{ |\varphi_p(\tilde{\mathbf{x}},\tilde{\mathbf{y}})|^2}{|\mathbf{r}_j^l(\mathbf{x},\mathbf{y},t)-\tilde{\mathbf{y}}|}
\;d\tilde{\mathbf{x}}d\tilde{\mathbf{y}}
\nonumber \\ && +A^l\sum_{s=1}^3 \left(\varphi_j^l(\mathbf{x},\mathbf{y}) \frac{1}{T} \int_0^T\left(\frac{\partial \mathbf{r}_j^l(\mathbf{x},\mathbf{y},t)}{\partial x_s}\cdot \frac{\partial \mathbf{r}_j^l(\mathbf{x},\mathbf{y},t)}{\partial x_s}\right.\right.\nonumber \\ &&+\left.\left.\frac{\partial \mathbf{r}_j^l(\mathbf{x},\mathbf{y},t)}{\partial y_s}\cdot \frac{\partial \mathbf{r}_j^l(\mathbf{x},\mathbf{y},t)}{\partial y_s}\right)\;dt\right)
\nonumber \\  && -m_e \varphi_j^l(\mathbf{x},\mathbf{y})\frac{1}{T}\int_0^T\frac{\partial \mathbf{r}_j^l(\mathbf{x},\mathbf{y},t)}{\partial t}\cdot \frac{\partial \mathbf{r}_j^l(\mathbf{x},\mathbf{y},t)}{\partial t}\;dt \nonumber \\ &&-E_{j}^l(\mathbf{y}) \varphi_j^l(\mathbf{x},\mathbf{y})=0, \text{ in } \Omega \times \Omega,
\end{eqnarray}

\item
$$\int_\Omega |\varphi_j^l(\mathbf{x},\mathbf{y})|^2\;dx=1,\; \forall l \in \{0,\ldots,n-1\},\; j \in \{-l,\ldots,0,\ldots,l\},\; \mathbf{y} \in \Omega,$$

\item
\begin{eqnarray}
&&-\gamma_p \sum_{s=1}^3  \left(\frac{\partial^2 \varphi_p(\mathbf{x},\mathbf{y})}{\partial x_s^2}+ \frac{\partial^2 \varphi_p(\mathbf{x},\mathbf{y})}{\partial y_s^2}\right)
\nonumber \\ && +\varphi_p(\mathbf{x},\mathbf{y}) \sum_{l=0}^{n-1}  \frac{Ke^2 Z}{T}\int_0^T \int_\Omega \int_\Omega
\frac{|\varphi_j^l(\tilde{\mathbf{x}},\tilde{\mathbf{y}})|^2}
{|\mathbf{y}-\mathbf{r}_j^l(\tilde{\mathbf{x}},\tilde{\mathbf{y}},t)|}\;d\tilde{\mathbf{x}}d\tilde{\mathbf{y}}dt
\nonumber \\ && - Ke^2Z^2 \;\varphi_p(\mathbf{x},\mathbf{y})\int_\Omega\int_\Omega\frac{ |\varphi_p(\tilde{\mathbf{x}},\tilde{\mathbf{y}})|^2}{|\mathbf{y}-\tilde{\mathbf{y}}|}
\;d\tilde{\mathbf{x}}d\tilde{\mathbf{y}}
\nonumber \\ &&  \nonumber \\ &&-E_p(\mathbf{y}) \varphi_p(\mathbf{x},\mathbf{y})=0, \text{ in } \Omega\times \Omega.
\end{eqnarray}

\item
$$\int_\Omega |\varphi_p(\mathbf{x},\mathbf{y})|^2\;d\mathbf{x}=1,\; \forall \mathbf{y} \in \Omega.$$

\end{enumerate}
\section{A note on the Entropy concept}

First define, for a wave function in a non-relativistic free particle context, for a motion developing on a time interval $[0,T],$ $$W(E)=\frac{1}{T}\int\int_{\Omega_E}|\phi(x)|^2\;dx\;dt$$ where
$$\Omega_{E}=\{ (x,t) \in \Omega \times [0,T]\;:\; E(x,t) \leq E\}.$$

At this point we define the entropy $S$ by,
\begin{eqnarray}S(E)&=&-\int_{E_0}^E W(\hat{E}) \ln(W(\hat{E})) \;d \hat{E},
\end{eqnarray}
where $E(x,t)$ will be specified in the next lines.

We define also the temperature $\hat{T}=\hat{T}(E)$ through the relation
$$\frac{dS(E)}{dE}=\frac{1}{\hat{T}}=-W(E)\ln W(E),$$ where we must emphasize the dependence $\hat{T}=\hat{T}(E).$

In a free particle context, we assume
\begin{eqnarray}E=E(x,t)&=&m_e|\phi(x)|^2\frac{\partial \mathbf{r}(x,t)}{\partial t}\cdot \frac{\partial \mathbf{r}(x,t)}{\partial t}
\nonumber \\ &=& \hat{E}_q-\mu |\phi(x)|^2,\end{eqnarray}
where here $\mu$ is such that $$\int_\Omega |\phi(x)|^2\;dx=1.$$

Also,
\begin{eqnarray}
\hat{E}_q(x,t)&=& -\gamma \sum_{k=1}^3 \frac{\partial^2 (\phi \mathbf{n})}{\partial x_k^2}\cdot  (\phi \mathbf{n}) \nonumber \\ &=& E_1(x,t) \mathbf{n} \cdot \mathbf{n} \nonumber \\
&=& E_1(x,t),
\end{eqnarray}
where $E_1(x,t)$ is the Lagrange multiplier such that $$\mathbf{n} \cdot \mathbf{n}=1, \text{ in } \Omega \times [0,T].$$
Summarizing,
$$E(x,t)=E_1(x,t)-\mu|\phi(x)|^2.$$

Finally,
\begin{eqnarray}d S(E)&=&-W(E)\ln(W(E))\;dE \nonumber \\ &=& \frac{dE}{\hat{T}}.
\end{eqnarray}
Hence, $$dS(E(x,t))=\frac{dE_1(x,t)}{\hat{T}}-\frac{d(\mu  |\phi(x)|^2)}{\hat{T}}.$$
\section{About modeling a chemical reaction}
Let us consider a volume $\Omega \subset \mathbb{R}^3$ and a possible  chemical reaction in $\Omega$,  in which $\alpha_1$ units of mass of a solid substance type $1$ reacts with $\alpha_2$ units of mass of a liquid
of type $2$ to produce 1 (one) unit of mass of a gaseous substance of type $3$, that is
$$\alpha_1+\alpha_2=1.$$

For such a system, we define
$$w_i(x,t)=\left(\frac{\rho_i(x,t) }{\sum_{k=1}^3\rho_k(x,t)}\right)^{\beta_i(x,t)},$$
where $\rho_k(x,t)$ denotes the point wise density of substance of type $k$, and $$\beta_k(x,t)=\beta_k(P(x,t),\hat{T}(x,t),\rho(x,t))$$ must be obtained
experimentally.

We define also, for such a system,
$$S(w)=-\sum_{k=1}^3\int_0^T\int_\Omega w_k(x,t) \ln(w_k(x,t)) \;dx\;dt.$$
\begin{obe} The functions $\beta_k(\hat{T},P,\rho)$ must be obtained such that the direction of the chemical reaction is properly modeled for the
concerning point-wise values of $\hat{T}(x,t),P(x,t), \rho(x,t).$
\end{obe}
\subsection{About the variational formulation modeling such a chemical reaction}

We define the problem of finding a critical point of a functional $J(\mathbf{r},\phi,E,\lambda,\mathbf{n},\hat{T},\mathbf{u})$, that is the problem of finding a solution for the
equation,
$$\delta J(\mathbf{r},\phi,E,\lambda,\mathbf{n},\hat{T},\mathbf{u})=\mathbf{0},$$
where
$J:U \times V_1 \times V_2 \times V_3 \times V_4 \times V_5 \times V_6\rightarrow \mathbb{R} $ will be specified in the next lines. In this model we consider the substance $s$ comprised by atoms of type $s$ with $Z_s$ electrons, protons and neutrons,
where $Z_s=\sum_{l=0}^{n_s}(2l+1)$ and  $n_s$ is the number of electronic layers for each atom of type $s$. We assume each layer $l$ has initially $2l+1$ electrons and the possible
molecular arrangements are obtained from the system motion and behavior, locally and as a whole.

We also define,
$$U=\left\{\mathbf{r}=\{(\mathbf{r}_s)_j^l\} \in W^{1,2}\left(\Omega \times \Omega \times [0,T];\mathbb{R}^{\sum_{s=1}^3 3Z_s}\right)\;:\; (\mathbf{r}_s)_j^l=\mathbf{x}
\text{ on } \Gamma_0, \text{ on }[0,T]\right\},$$
\begin{eqnarray}V_1&=&\left\{\phi=\{(\phi_s)_j^l,(\phi_p)_s\} \in W^{1,2}\left(\Omega \times \Omega \times[0,T];\mathbb{C}^{\left(\sum_{s=1}^3 Z_s\right)+ 3 }\right)\right.
\nonumber \\ &&\;:\;\left.
(\phi_s)_j^l=(\phi_p)_s=0, \text{ on }  \Gamma_1 \text{ on } [0,T]\right\},\end{eqnarray}
where $$\Gamma_0,\; \Gamma_1 \subset \partial(\Omega \times \Omega),$$

$$V_2=\mathbb{R}^{\left(\sum_{s=1}^3  Z_s\right) + 3},$$
$$V_3=L^2\left(\Omega \times [0,T];\mathbb{R}^{5+ \left(\sum_{s=1}^3  2 Z_s\right)}\right),$$
$$V_4=L_2([0,T]; \mathbb{R}^3),$$
$$V_5=W^{2,2}(\Omega \times [0,T];\mathbb{R}^4)$$
and $$V_6=L^2\left(\Omega \times [0,T]; \mathbb{R}^{\sum_{s=1}^3  Z_s}\right).$$
Finally, the functional $J$ is expressed as

$$J=J_1+J_2+J_3+J_4+J_5+J_6+J_7,$$
where
\begin{eqnarray}
 J_1(\mathbf{r},\phi)&=&
-\sum_{s=1}^3 \sum_{l=0}^{n_s-1} \sum_{j=-l}^l \int_0^T\int_\Omega \int_\Omega |(\phi_s)_j^l|^2\frac{\partial (\mathbf{r}_s)_j^l}{\partial t}
\cdot \frac{\partial (\mathbf{r}_s)_j^l}{\partial t}\;d \mathbf{x}\;d \mathbf{y}\;dt\nonumber \\ &&-
\sum_{s=1}^3\int_0^T\int_\Omega\int_\Omega |(\phi_p)_s|^2\; \frac{\partial (\mathbf{r}_p)_s}{\partial t} \cdot \frac{\partial (\mathbf{r}_p)_s}{\partial t}
\;d \mathbf{x}\;d \mathbf{y}\;dt, \end{eqnarray}
\begin{eqnarray}
&&J_2(\mathbf{r},\phi)\nonumber \\ &=& \sum_{s=1}^3\sum_{q=1}^3\sum_{l=0}^{n_s-1} \sum_{l_1=0}^{n_q-1} \sum_{k=-l_1}^{l_1}\nonumber \\ &&\left(Ke^2\int_0^T\int_\Omega\int_\Omega\int_\Omega\int_\Omega
 \frac{|(\phi_s)_j^l(\mathbf{x},\mathbf{y},t)|^2|(\phi_q)_k^{l_1}(\tilde{\mathbf{x}},\tilde{\mathbf{y}},t)|^2}{|(\mathbf{r}_s)_j^l(\mathbf{x},\mathbf{y},t)-
(\mathbf{r}_q)_k^{l_1}(\tilde{\mathbf{x}},\tilde{\mathbf{y}},t)|}\;d \mathbf{x}\;d \mathbf{y}\;d \tilde{\mathbf{x}}\;d \tilde{\mathbf{y}}\;dt \right)\nonumber \\ &&
-\sum_{s=1}^3\sum_{q=1}^3\sum_{l=0}^{n_s-1} Ke^2\int_0^T\int_\Omega\int_\Omega\int_\Omega\int_\Omega
 \frac{|(\phi_s)_j^l(\mathbf{x},\mathbf{y},t)|^2|(\phi_p)_q(\tilde{\mathbf{x}},\tilde{\mathbf{y}},t)|^2}{|(\mathbf{r}_s)_j^l(\mathbf{x},\mathbf{y},t)
-(\mathbf{r}_p)_q(\tilde{\mathbf{x}},\tilde{\mathbf{y}},t)|}\;d \mathbf{x}\;d \mathbf{y}\;d \tilde{\mathbf{x}}\;d \tilde{\mathbf{y}}\;dt \nonumber \\ &&+
\sum_{s=1}^3\sum_{q=1}^3 Ke^2\int_0^T\int_\Omega\int_\Omega\int_\Omega\int_\Omega
 \frac{|(\phi_p)_s(\mathbf{x},\mathbf{y},t)|^2|(\phi_p)_q(\tilde{\mathbf{x}},\tilde{\mathbf{y}},t)|^2}{|(\mathbf{r}_p)_s(\mathbf{x},\mathbf{y},t)
-(\mathbf{r}_p)_q(\tilde{\mathbf{x}},\tilde{\mathbf{y}},t)|}\;d \mathbf{x}\;d \mathbf{y}\;d \tilde{\mathbf{x}}\;d \tilde{\mathbf{y}}\;dt,
\end{eqnarray}
\begin{eqnarray}
&&J_3(\mathbf{r},\phi,\mathbf{n})\nonumber \\ &=& \sum_{s=1}^3 \sum_{l=0}^{n_s-1} \sum_{j=-l^l}(\gamma_s)_j^l \int_0^T \int_\Omega \int_\Omega
\sum_{k=1}^3\left( \frac{\partial [(\phi_s)_j^l (\mathbf{n}_s)_j^l]}{\partial x_k} \cdot \frac{\partial [(\phi_s)_j^l (\mathbf{n}_s)_j^l]}{\partial x_k}
\right.\nonumber \\ &&\left.+\frac{\partial [(\phi_s)_j^l (\mathbf{n}_s)_j^l]}{\partial y_k} \cdot \frac{\partial [(\phi_s)_j^l (\mathbf{n}_s)_j^l]}{\partial y_k}\right)\;d\mathbf{x}\;d\mathbf{y}\;dt \nonumber \\ &&
+\sum_{s=1}^3(\gamma_p)_s \int_0^T \int_\Omega \int_\Omega \sum_{k=1}^3\left( \frac{\partial [(\phi_s)_j^l (\mathbf{n}_s)_j^l]}{\partial x_k} \cdot \frac{\partial [(\phi_s)_j^l (\mathbf{n}_s)_j^l]}{\partial x_k}
\right.\nonumber \\ && \left.+\frac{\partial [(\phi_s)_j^l (\mathbf{n}_s)_j^l]}{\partial y_k} \cdot \frac{\partial [(\phi_s)_j^l (\mathbf{n}_s)_j^l]}{\partial y_k}\right)\;d\mathbf{x}\;d\mathbf{y}\;dt.
\end{eqnarray}
Also,
\begin{eqnarray}
&&J_4(\mathbf{r},\phi,E,\lambda,\hat{T},\mathbf{u})\nonumber \\ &=& \frac{1}{2}\int_0^T\int_\Omega H_{ijkl}(\rho_1) \epsilon_{ij}(\hat{\mathbf{u}})  \epsilon_{kl}(\hat{\mathbf{u}})
\;(1-\chi_{\rho_f})\;d\mathbf{x}\;dt
\nonumber \\ &&-\int_0^T\int_\Omega f_i\hat{u}_i\;(1-\chi_{\rho_f})\;d\mathbf{x}\;dt-\int_0^T\int_{\Gamma_t} \hat{f}_i\hat{u}_i\;(1-\chi_{\rho_f})\;d\Gamma\;dt
\nonumber \\ && +\sum_{k=1}^3\int_0^T \int_\Omega \lambda_k\left(\frac{\partial (\rho u_k)}{\partial t}
+\sum_{j=1}^3\frac{\partial [(\rho u_k) u_j]}{\partial x_j}-\sum_{j=1}^3\frac{\partial \tau_{jk}}{\partial x_j}+\frac{\partial P}{\partial x_k}-g_k\right)\;\chi_{\rho_f}\;d\mathbf{x}\;dt
\nonumber \\ &&+\int_0^T\int_\Omega\lambda_4 \left(\frac{\partial \rho}{\partial t}+\sum_{j=1}^3\frac{\partial (\rho u_j)}{\partial x_j}\right)\;\chi_{\rho_f}\;d\mathbf{x}\;dt
\nonumber \\ &&+\int_0^T\int_\Omega \lambda_5\left( \frac{ \partial E_n}{\partial t}+\sum_{j=1}^3\frac{\partial}{\partial x_j}\left( u_j E_n-\sum_{k=1}^3u_k \tau_{jk}-q_j\right)\right)\chi_{\rho_f}
\;d\mathbf{x}\;dt, \end{eqnarray}
where $\Gamma_t \subset \partial \Omega$ and for appropriate constants $\hat{K}>0, \; R>0$ and $K_B>0$, we have
$$\mathbf{q}=\hat{K} \nabla \hat{T},$$
$$P= \rho_3 R\hat{T},$$
$$\rho_f=\rho_2+\rho_3,$$
$$\tau_{ij}=-\frac{\rho}{m_f} K_B\hat{T} \delta_{ij}-\frac{2}{3}\mu(\rho,\hat{T},P)\sum_{k=1}^3\frac{\partial u_k}{\partial x_k} \delta_{ij}
+\mu(\rho,\hat{T},P)\left(\frac{\partial u_i}{\partial x_j}+\frac{\partial u_j}{\partial x_i}\right),$$
and
%$$E=\frac{3}{2}\frac{m_f}{\rho_f}K_B T+\frac{1}{2}\rho_f |\mathbf{u}|^2.$$
\begin{equation} E_n(x,t)=\left\{
\begin{array}{ll}
 \frac{3}{2}\frac{m_f}{\rho}K_B \hat{T}+\frac{1}{2}\rho_f |\mathbf{u}|^2,& \text{ if } \rho_f(x,t)\neq 0,
 \\
0, &\text{ if }  \rho_f(x,t)=0.\end{array} \right.\end{equation}
where $$m_f(t)=\int_\Omega \rho(x,t)\chi_{\rho_f}\;d\mathbf{x}.$$

Moreover, we define the strain tensor (for the solid type $1$), by $$\epsilon(\hat{\mathbf{u}})=\{\epsilon_{ij}(\hat{\mathbf{u}})\}=\left\{ \frac{\hat{u}_{i,j}+\hat{u}_{j,i}}{2}\right\}.$$

 On the other hand, $$\{H_{ijkl}(\rho_1)\}$$ is a positive definite tensor which represents the stiffness matrix for the solid type $1$, which is assumed to depend linearly
on $\rho_1.$ Also, $f \in L^2(\Omega;\mathbb{R}^3)$ and $\hat{f} \in L^2(\partial \Omega;\mathbb{R}^3)$ are external loads effectively acting on the solid type $1$ only where
$\chi_{\rho_f}=0.$

\begin{obe} We   assume the concerning tensor is such that  $$H_{ijkl}(\rho_1(x,t))\approx 0$$ if $$\rho_1(x,t)\approx 0,$$
and expect, in an appropriate sense, at least approximately
$$\chi_{\rho_1}\approx 1-\chi_{\rho_f}.$$

Here, generically,
\begin{equation} \chi_{\rho_f}(x,t)=\left\{
\begin{array}{ll}
1,& \text{ if } \rho_f(x,t)>0,
 \\
0, &\text{ if }  \rho_f(x,t)=0.\end{array} \right.\end{equation}

\end{obe}

Furthermore,  denoting the initial mass of the substance type $s$ by $(m_0)_s$, we have
\begin{eqnarray}
m_s(t)&=&(m_0)_s
-\int_0^t\int_{\partial \Omega}  \rho_s(\mathbf{x},\hat{t}) \mathbf{u} \cdot \mathbf{n}\; d \Gamma\;d \hat{t}
\nonumber \\ && -\int_0^t\int_{\partial \Omega} \alpha_s \rho_3(\mathbf{x},\hat{t}) \mathbf{u} \cdot \mathbf{n}\; d \Gamma\;d \hat{t}
\end{eqnarray}
$\forall s \in \{1,2\}$, where here $\mathbf{n}=(n_1,n_2,n_3)$ denotes the outward normal field to $\partial \Omega$ and $m_s(t)$ denotes the mass of substance type $s$ at the time $t$. We emphasize to have assumed $(m_0)_3=0$ and the substance type $3$ may
only leave the system represented by $\Omega$ (not enter it).

Considering such assumptions and statements, we define
\begin{eqnarray}
&&J_5(\mathbf{r},\phi,E,\lambda,\mathbf{u})\nonumber \\ &=& -\frac{1}{2} \sum_{s=1}^3\sum_{l=0}^{n_s-1}\sum_{j=-l}^l
\int_0^T\int_\Omega (E_s)_j^l(\mathbf{y},t) \left(\int_\Omega |(\phi_s)_j^l(\mathbf{x},\mathbf{y},t)|^2\;d\mathbf{x}-(m_e)_s(t)\right)\;d\mathbf{y}\;dt
\nonumber \\ && -\frac{1}{2} \sum_{s=1}^3
\int_0^T\int_\Omega (E_s)_p(\mathbf{y},t) \left(\int_\Omega |(\phi_s)_p(\mathbf{x},\mathbf{y},t)|^2\;d\mathbf{x}-\frac{(m_p+m_n)}{m_e} (m_e)_s(t) Z_s\right)\;d\mathbf{y}\;dt
\nonumber \\ && +\int_0^T\sum_{s=1}^2 \lambda_{5+s}(t)\left(m_s(t)-(m_0)_s
+\int_0^t\int_{\partial \Omega}  \rho_s(\mathbf{x},\hat{t}) \mathbf{u} \cdot \mathbf{n}\; d \Gamma\;d \hat{t}
\right.\nonumber \\ && \left.+\int_0^t\int_{\partial \Omega} \alpha_s \rho_3(\mathbf{x},\hat{t}) \mathbf{u} \cdot \mathbf{n}\; d \Gamma\;d \hat{t}\right)\;dt
\nonumber \\ && +\int_0^T \lambda_9(t)\left(m_1(t)+m_2(t)+m_3(t)\right.
\nonumber \\ &&\left.-\left((m_0)_1+(m_0)_2-\sum_{s=1}^3\int_0^t\int_{\partial \Omega}  \rho_s(\mathbf{x},\hat{t})
 \mathbf{u} \cdot \mathbf{n}\; d \Gamma\;d \hat{t}\right)\right)\; dt,
\end{eqnarray}
where $(m_e)_s(t)$ is such that $$Z_s(m_e)_s(t)+Z_s \frac{m_p+m_n}{m_e}(m_e)_s(t)=m_s(t),\; \forall s \in \{1,2,3\}.$$
Here $m_e,m_p,m_n$ denotes the mass of a single electron, proton and neutron, respectively.

We also define
\begin{eqnarray}
J_6(\mathbf{r},\lambda,\mathbf{n})&=& \sum_{s=1}^3\sum_{l=0}^{n_s-1}\sum_{j=-l}^l \int_0^T \int_\Omega (\lambda_{7}^s)_j^l
((\mathbf{n}_s)_j^l\cdot (\mathbf{n}_s)_j^l-1)\;d\mathbf{x}dt \nonumber \\ &&
+\sum_{s=1}^3\sum_{l=0}^{n_s-1}\sum_{j=-l}^l\int_0^T \int_\Omega (\lambda_{8}^s)_j^l
(\mathbf{n}_s)_j^l\cdot \frac{\partial (\mathbf{r}_s)_j^l}{\partial t}\;d\mathbf{x}dt
\end{eqnarray}
and
$$
J_7(\phi,\hat{T})= -S(w),$$
where \begin{eqnarray}\rho_s(\mathbf{x},t)&=&\sum_{l=0}^{n_s-1}\sum_{j=-l}^l \int_\Omega |(\phi_s)_j^l(\mathbf{x},\mathbf{y},t)|^2\;d\mathbf{y}
\nonumber \\ &&+\int_\Omega |(\phi_p)_s(\mathbf{x},\mathbf{y},t)|^2\;d \mathbf{y},
\end{eqnarray}
and $$m_s(t)=\int_\Omega \rho_s(\mathbf{x},t)\;d\mathbf{x}.$$

Moreover,
$$(\mathbf{r}_s)_j^l(\mathbf{x},\mathbf{y},t)=\mathbf{r}_s(\mathbf{x},t)+\hat{\mathbf{r}}_s(\mathbf{x},t)
+(\tilde{\mathbf{r}}_{s})_j^l(\mathbf{x},\mathbf{y},t)\approx \mathbf{x},$$
where $$\mathbf{r}_s(\mathbf{x},t)=\mathbf{x}, \text{ in } \Omega.$$

Also,  $$\hat{\mathbf{r}}_s(\mathbf{x},t)=\mathbf{0},\text{ if } s=2,3,$$ and $$\hat{\mathbf{r}}_1(\mathbf{x},t)=\hat{\mathbf{u}}(\mathbf{x},t)=(\hat{u}_1(\mathbf{x},t),\hat{u}_2(\mathbf{x},t),\hat{u}_3(\mathbf{x},t))$$ refers to the displacement field for the solid part,
and $$\mathbf{u}=(u_1,u_2,u_3)$$ is the velocity field for the fluid part.

Furthermore,
$$(\mathbf{r}_p)_s(\mathbf{x},\mathbf{y},t)=\mathbf{y}+(\hat{\mathbf{r}}_p)_s(\mathbf{x},\mathbf{y},t) \approx \mathbf{y}.$$
%\begin{equation} (\mathbf{r}_{1})_s(\mathbf{x},t)=\left\{
%\begin{array}{ll}
% \mathbf{x},& \text{ if } s=1,
% \\
%\mathbf{r}_1(\mathbf{x},t), &\text{ if }  s=2,3.\end{array} \right.\end{equation}
\subsection{The final variational formulation}
This previous variational formulation may be useful in a nano-technology context, for example.

However, since it is a multi-scale one, it is of difficult computation. So, with such statements in mind, we shall propose a final
macroscopic version for such a model, which we shall denote by $\tilde{J}$.

Concerning an analogy relating the previous formulation, in the next lines we set,
$$(\mathbf{r}_s)_j^l(\mathbf{x},\mathbf{y},t)= \mathbf{x},\; \text{ for } s=2,3$$
which translates into $$\mathbf{r}_s(\mathbf{x},t)=\mathbf{x},\; \text{ for } s=2,3,$$
and
$$\mathbf{r}_1(\mathbf{x},t)=\mathbf{x}+\hat{\mathbf{u}}(\mathbf{x},t)=\mathbf{x}+(\hat{u}_1(\mathbf{x},t),\hat{u}_2(\mathbf{x},t),\hat{u}_3(\mathbf{x},t)).$$

Moreover, $$\mathbf{u}=(u_1,u_2,u_3)$$ is the velocity field for the fluid part.

Finally, we also set $$(\mathbf{n_s})_j^l=\mathbf{k}_0$$ for an appropriate unit constant vector $\mathbf{k}_0 \in \mathbb{R}^3.$

Concerning the new proposed formulation, we define,
$$\tilde{J}=\tilde{J}_1+\tilde{J}_2+\tilde{J}_3+\tilde{J}_4+\tilde{J}_5,$$
where
\begin{eqnarray}
 \tilde{J}_1(\mathbf{r},\phi)&=&
-\sum_{s=1}^3 \int_0^T\int_\Omega  |\phi_1|^2\frac{\partial \mathbf{r}_1}{\partial t}
\cdot \frac{\partial \mathbf{r}_1}{\partial t}\;d \mathbf{x}\;\;dt, \end{eqnarray}
\begin{eqnarray}
&&\tilde{J}_2(\mathbf{r},\phi)\nonumber \\ &=& \sum_{s=1}^3\sum_{q=1}^3K\int_0^T\int_\Omega\int_\Omega
 \frac{|\phi_s(\mathbf{x},t)|^2|\phi_q(\tilde{\mathbf{x}},t)|^2}{|\mathbf{r}_s(\mathbf{x},t)-
\mathbf{r}_q(\tilde{\mathbf{x}},t)|}\;d \mathbf{x}\;d \tilde{\mathbf{x}}\;dt
\end{eqnarray}
\begin{eqnarray}
&&\tilde{J}_3(\phi)\nonumber \\ &=& \sum_{s=1}^3 \gamma_s\int_0^T \int_\Omega
\sum_{k=1}^3\left( \frac{\partial \phi_s}{\partial x_k} \; \frac{\partial \phi_s}{\partial x_k}
\right)\;d\mathbf{x}\;dt.
\end{eqnarray}
Also,
\begin{eqnarray}
&&\tilde{J}_4(\mathbf{r},\phi,E,\lambda,\hat{T},\mathbf{u})\nonumber \\ &=& \frac{1}{2}\int_0^T\int_\Omega H_{ijkl}(\rho_1) \epsilon_{ij}(\hat{\mathbf{u}})  \epsilon_{kl}(\hat{\mathbf{u}})
\;(1-\chi_{\rho_f})\;d\mathbf{x}\;dt
\nonumber \\ &&-\int_0^T\int_\Omega f_i\hat{u}_i\;(1-\chi_{\rho_f})\;d\mathbf{x}\;dt-\int_0^T\int_{\Gamma_t} \hat{f}_i\hat{u}_i\;(1-\chi_{\rho_f})\;d\Gamma\;dt
\nonumber \\ && +\sum_{k=1}^3\int_0^T \int_\Omega \lambda_k\left(\frac{\partial (\rho u_k)}{\partial t}
+\sum_{j=1}^3\frac{\partial [(\rho u_k) u_j]}{\partial x_j}-\sum_{j=1}^3\frac{\partial \tau_{jk}}{\partial x_j}+\frac{\partial P}{\partial x_k}-g_k\right)\;\chi_{\rho_f}\;d\mathbf{x}\;dt
\nonumber \\ &&+\int_0^T\int_\Omega\lambda_4 \left(\frac{\partial \rho}{\partial t}+\sum_{j=1}^3\frac{\partial (\rho u_j)}{\partial x_j}\right)\;\chi_{\rho_f}\;d\mathbf{x}\;dt
\nonumber \\ &&+\int_0^T\int_\Omega \lambda_5\left( \frac{ \partial E_n}{\partial t}+\sum_{j=1}^3\frac{\partial}{\partial x_j}\left( u_j E_n-\sum_{k=1}^3u_k \tau_{jk}-q_j\right)\right)\chi_{\rho_f}
\;d\mathbf{x}\;dt, \end{eqnarray}
where again $$\mathbf{q}=\hat{K} \nabla \hat{T},$$
$$P= \rho_3 R\hat{T},$$
$$\rho_f=\rho_2+\rho_3,$$
$$\tau_{ij}=-\frac{\rho}{m_f} K_BT \delta_{ij}-\frac{2}{3}\mu(\rho,\hat{T},P)\sum_{k=1}^3\frac{\partial u_k}{\partial x_k} \delta_{ij}
+\mu(\rho,\hat{T},P)\left(\frac{\partial u_i}{\partial x_j}+\frac{\partial u_j}{\partial x_i}\right),$$
and
%$$E=\frac{3}{2}\frac{m_f}{\rho_f}K_B T+\frac{1}{2}\rho_f |\mathbf{u}|^2.$$
\begin{equation} E_n(x,t)=\left\{
\begin{array}{ll}
 \frac{3}{2}\frac{m_f}{\rho}K_B \hat{T}+\frac{1}{2}\rho_f |\mathbf{u}|^2,& \text{ if } \rho_f(x,t)\neq 0,
 \\
0, &\text{ if }  \rho_f(x,t)=0.\end{array} \right.\end{equation}
where $$m_f(t)=\int_\Omega \rho(x,t)\chi_{\rho_f}\;d\mathbf{x}.$$

Here also again, generically
\begin{equation} \chi_{\rho_f}(x,t)=\left\{
\begin{array}{ll}
1,& \text{ if } \rho_f(x,t)>0,
 \\
0, &\text{ if }  \rho_f(x,t)=0.\end{array} \right.\end{equation}

Finally,
\begin{eqnarray}
&&\tilde{J}_5(\mathbf{r},\phi,\lambda,\hat{T},\mathbf{u})\nonumber \\ &=& \int_0^T\sum_{s=1}^2 \lambda_{5+s}(t)\left(m_s(t)-(m_0)_s
+\int_0^t\int_{\partial \Omega}  \rho_s(\mathbf{x},\hat{t}) \mathbf{u} \cdot \mathbf{n}\; d \Gamma\;d \hat{t}
\right.\nonumber \\ && \left.+\int_0^t\int_{\partial \Omega} \alpha_s \rho_3(\mathbf{x},\hat{t}) \mathbf{u} \cdot \mathbf{n}\; d \Gamma\;d \hat{t}\right)\;dt
\nonumber \\&&+\int_0^T \lambda_8(t)\left(m_1(t)+m_2(t)+m_3(t)\right.\nonumber \\ &&\left.-\left((m_0)_1+(m_0)_2-\sum_{s=1}^3\int_0^t\int_{\partial \Omega}  \rho_s(\mathbf{x},\hat{t})
 \mathbf{u} \cdot \mathbf{n}\; d \Gamma\;d \hat{t}\right)\right)\; dt  \nonumber \\ &&-S(w),
\end{eqnarray}
where \begin{eqnarray}\rho_s(\mathbf{x},t)= |(\phi_s)(\mathbf{x},t)|^2,
\end{eqnarray}
 $$m_s(t)=\int_\Omega \rho_s(\mathbf{x},t)\;d\mathbf{x},\; \forall s \in \{1,2,3\}$$
and $$\rho(\mathbf{x},t)= \sum_{s=1}^3 \rho_s(\mathbf{x},t).$$

\section{ A note on the Spin operator}
We finish this article with a result about the Spin operator in a relativistic context.

Consider a wave function $\phi(\mathbf{r})$ related to the scalar density field of a particle  with position field given by
$$\mathbf{r}:\Omega \times [0,T] \rightarrow \mathbb{R}^3.$$

Observe that in a special relativity context the field of velocity $$\frac{\partial \mathbf{r}(\mathbf{x},t)}{\partial t}$$ induces a Lorentz type transformation
concerning an observer  at $(0,0,0) \in \mathbb{R}^3.$ So the corresponding transform of the vector
$$\mathbf{r}(\mathbf{x},t)=(ct,X_1(\mathbf{x},t),X_2(\mathbf{x},t),X_3(\mathbf{x},t))$$ will be the vector
$$(ct',X_1',X_2',X_3'),$$ where
$$X_j'=\left(\frac{1}{\sqrt{1-\frac{v^2}{c^2}}}-1 \right)\left(\frac{\mathbf{r}_1\cdot \mathbf{v}}{v^2}\right)v_j-\frac{tv_j}{\sqrt{1-\frac{v^2}{c^2}}}+X_j(\mathbf{x},t),$$
$\forall j \in \{1,2,3\},$
and
$$t'=\frac{1}{\sqrt{1-\frac{v^2}{c^2}}} \left(t-\frac{\mathbf{r}_1\cdot \mathbf{v}}{c^2}\right),$$
where
$$\mathbf{r}_1(\mathbf{x},t)=(X_1(\mathbf{x},t),X_2(\mathbf{x},t),X_3(\mathbf{x},t)),$$
 $$\mathbf{v}=\frac{\partial \mathbf{r}_1(\mathbf{x},t)}{\partial t},$$
 and $$v=\sqrt{\left(\frac{\partial X_1(\mathbf{x},t)}{\partial t}\right)^2+\left(\frac{\partial X_2(\mathbf{x},t)}{\partial t}\right)^2+
 \left(\frac{\partial X_3(\mathbf{x},t)}{\partial t}\right)^2}.$$

 We assume there exists a function such that $\varphi$ $$\phi(\mathbf{r}(\mathbf{x},t))=\varphi(X_1',X_2',X_3',t').$$

 At this point we shall define the angular momentum operator.

 First, we consider a rotation about the $z$ axis, so that we define
 \begin{eqnarray}\mathbf{r}_\varepsilon(\mathbf{x},t) &=&(X_1(\mathbf{x},t),X_2(\mathbf{x},t),X_3(\mathbf{x},t))+ \varepsilon(- x_2,x_1,0)
 \nonumber \\ &=& ((X_1)_\varepsilon,(X_2)_\varepsilon,(X_3)_\varepsilon),\end{eqnarray}
 where
 $$(X_1)_\varepsilon =X_1(\mathbf{x},t)-\varepsilon x_2,$$
 $$(X_2)_\varepsilon =X_2(\mathbf{x},t)+\varepsilon x_1,$$
 $$(X_3)_\varepsilon =X_3(\mathbf{x},t),$$
 and also
 $$t_\varepsilon =t.$$

 In such a case, we have $$\mathbf{v}_\varepsilon=\frac{\partial \mathbf{r}_\varepsilon(\mathbf{x},t)}{\partial t}=\mathbf{v},$$ so that
 $$v_\varepsilon=v.$$

 Hence, we define the angular momentum coordinate $J_z(\varphi(\mathbf{X}',t'))$, by
 \begin{eqnarray}
  J_z(\varphi(\mathbf{X}',t'))=-i \hbar \frac{\partial \varphi(\mathbf{X}_\varepsilon',t_\varepsilon'))}{\partial \varepsilon}|_{\varepsilon=0},
  \end{eqnarray}
where
$$\mathbf{X}_\varepsilon'= \left(\frac{1}{\sqrt{1-\frac{v^2}{c^2}}}-1 \right)\left(\frac{\mathbf{r}_\varepsilon\cdot \mathbf{v}}{v^2}\right)\mathbf{v}-\frac{t\mathbf{v}}{\sqrt{1-\frac{v^2}{c^2}}}+\mathbf{r}_\varepsilon(\mathbf{x},t),$$
and
$$t_\varepsilon'=\frac{1}{\sqrt{1-\frac{v^2}{c^2}}} \left(t-\frac{\mathbf{r}_\varepsilon\cdot \mathbf{v}}{c^2}\right),$$
so that
\begin{eqnarray}
J_z(\varphi(\mathbf{X}',t'))&=&-i \hbar \sum_{j=1}^3\frac{\partial \varphi(\mathbf{X}_\varepsilon',t_\varepsilon'))}{\partial X_j} \frac{\partial (X_\varepsilon')_j}{\partial \varepsilon}|_{\varepsilon=0} \nonumber \\ &&-i \hbar\frac{\partial \varphi(\mathbf{X}_\varepsilon',t_\varepsilon'))}{\partial t} \frac{\partial t_\varepsilon'}{\partial \varepsilon}|_{\varepsilon=0}.
\end{eqnarray}

Observe that
$$\frac{\partial \mathbf{X}_\varepsilon'}{\partial \varepsilon}=
 \left(\frac{1}{\sqrt{1-\frac{v^2}{c^2}}}-1 \right)\left(\frac{(-x_2,x_1,0)\cdot \mathbf{v}}{v^2}\right)\mathbf{v} +(-x_2,x_1,0),$$
 and
 $$\frac{\partial t_\varepsilon'}{\partial \varepsilon}=
 -\frac{1}{\sqrt{1-\frac{v^2}{c^2}}}\frac{(-x_2,x_1,0) \cdot \mathbf{v}}{c^2}.$$

From such last results, we have
\begin{eqnarray}
J_z(\varphi(\mathbf{X}',t'))&=& -i\hbar \sum_{j=1}^3 \frac{\partial \varphi(\mathbf{X}',t')}{\partial X_j}
\left(\frac{1}{\sqrt{1-\frac{v^2}{c^2}}}-1 \right)\left(\frac{(-x_2,x_1,0)\cdot \mathbf{v}}{v^2}\right)v_j \nonumber \\ &&
+i\hbar \frac{\partial \varphi(\mathbf{X}',t')}{\partial t}\frac{1}{\sqrt{1-\frac{v^2}{c^2}}}\frac{(-x_2,x_1,0) \cdot \mathbf{v}}{c^2}\nonumber \\ &&
-i\hbar \left( -x_2\frac{\partial \varphi(\mathbf{X}',t')}{\partial X_1}+x_1\frac{\partial \varphi(\mathbf{X}',t')}{\partial X_2}\right)
\nonumber \\ &=& S_z(\varphi(\mathbf{X}',t'))+L_z(\varphi(\mathbf{X}',t')), \end{eqnarray}
where
$$L_z(\varphi(\mathbf{X}',t'))=-i\hbar \left( -x_2\frac{\partial \varphi(\mathbf{X}',t')}{\partial X_1}+x_1\frac{\partial \varphi(\mathbf{X}',t')}{\partial X_2}\right),$$
and
\begin{eqnarray}S_z(\varphi(\mathbf{X}',t'))&=&-i\hbar \sum_{j=1}^3 \frac{\partial \varphi(\mathbf{X}',t')}{\partial X_j}
\left(\frac{1}{\sqrt{1-\frac{v^2}{c^2}}}-1 \right)\left(\frac{(-x_2,x_1,0)\cdot \mathbf{v}}{v^2} \right)v_j
\nonumber \\ && +i\hbar \frac{\partial \varphi(\mathbf{X}',t')}{\partial t}\frac{1}{\sqrt{1-\frac{v^2}{c^2}}}\frac{(-x_2,x_1,0) \cdot \mathbf{v}}{c^2}.
\end{eqnarray}

Similarly we may obtain $L_x,L_y,S_x,S_y$.

Finally defining $$\mathbf{J}=\mathbf{S} + \mathbf{L},$$
where $$\mathbf{L}=(L_x,L_y,L_z)$$ and $$\mathbf{S}=(S_x,S_y,S_z),$$ we call $\mathbf{L}$ the orbital angular momentum operator and $\mathbf{S}$ the
spin one.

\section{Conclusion} In this article we have developed a variational formulation for the relativistic Klein-Gordon equation
by extending the standard classical mechanics energy to a more general functional.

We believe the results here presented may be applied and extended to other models in mechanics, including the quantum and relativistic approaches for
the study of atoms and molecules.

In one of the last sections, it has been presented  a first analysis including the presence of electromagnetic fields.

Finally, in the last section, we present a result about the Spin operator in a relativistic context.

\end{document}